\documentclass[twocolumn,aps,pra,groupedaddress,showpacs]{revtex4}
\usepackage{epsfig,amssymb,amsmath}
\def\comment#1{}\def\labell#1{\label{#1}}
\def\cm{{\cal M}}\def\cn{{\cal N}}\def\ce{{\cal E}}\def\ms{{\mathbb S}}
\begin{document}
\title{Minimum output entropy of bosonic channels:  a conjecture}
\author{Vittorio Giovannetti,$^1$\footnote{Now with NEST-INFM \& Scuola Normale Superiore,
Piazza dei Cavalieri 7, I-56126, Pisa, Italy.} Saikat Guha,$^1$ Seth Lloyd,$^{1,2}$
Lorenzo Maccone,$^1$\footnote{Now with QUIT - Quantum Information Theory Group
Dipartimento di Fisica ``A. Volta''
Universita' di Pavia, via A. Bassi 6
I-27100, Pavia, Italy.} and Jeffrey H.
  Shapiro,$^1$}\affiliation{$^1$Massachusetts
Institute of Technology
  -- Research Laboratory of Electronics\\$^2$Massachusetts Institute
  of Technology -- Department of Mechanical Engineering\\ 77
  Massachusetts Avenue, Cambridge, MA 02139, USA}

\begin{abstract}
The von Neumann entropy at the output of a bosonic channel with thermal
noise is analyzed.  Coherent-state inputs are conjectured to minimize this
output entropy. Physical and mathematical evidence in support of the
conjecture is provided.  A stronger conjecture---that output states
resulting from coherent-state inputs majorize the output states from
other inputs---is also discussed.
\end{abstract}
\pacs{03.67.Hk,03.67.-a,03.65.Db,42.50.-p} \maketitle

A quantum channel can be characterized by a completely positive (CP)
linear super-operator on the Hilbert space of the information
carrier.  In general this evolution is not unitary, so that a pure
state loses some coherence in its transit through the channel.
Various measures of a channel's ability to preserve the coherence of its
input state have been introduced. Arguably, the most
useful of these are the information capacities~\cite{chuang,shor,asomov}.
In contrast, we will focus on the minimum von Neumann
entropy $\ms$ at the channel output. This quantity is related to the
minimum amount of noise implicit in the channel: when the channel input
is a pure state, $\ms$ quantifies the minimum uncertainty occurring in
the resulting channel output.  More precisely, the output entropy
associated with a pure state measures the entanglement that such a state
establishes with the environment during the communication.  Because the
state of the environment is not accessible, this entanglement is
responsible for the loss of quantum coherence and hence for the
injection of noise into the channel output.  Low values of
entanglement, i.e., of $\ms$, then correspond to low-noise communication. 
Furthermore, the study of $\ms$ yields important information about
channel capacities. In particular, an upper bound on the classical
capacity derives from a lower bound on the output entropy of
multiple channel uses, see, e.g., \cite{ruskai,congettura}.  Finally,
the additivity of the minimum entropy implies the additivity of the
classical capacity and of the entanglement of formation~\cite{shorequiv}.

Our study of minimum output entropy will be restricted to 
bosonic channels in which the electromagnetic field, used as
the information carrier, interacts with a thermal-like noise source.  For
these channels we analyze the following conjecture: that minimum output
entropy is achieved when the channel input is a coherent state.  In what
follows the rationale and some physical justification for this conjecture
are presented.   We also consider a stronger conjecture---that output
states resulting from coherent-state inputs majorize the output states
from other inputs---which, if true, would imply the minimum output entropy
conjecture. (Note that the minimum output entropy problem was previously
treated in \cite{hall}, which reported some of the results that we will
discuss.)  Additional supporting evidence for the
minimum-entropy conjecture appears in our companion
paper~\cite{renyinostro}, where we show that the integer-order R\'enyi
entropies and the Wehrl entropy at the output of the bosonic channel are
minimized when the channel input is a coherent state.

In Sec.~\ref{s:nm} we present CP-maps for the two bosonic channels that 
will be considered in this paper.  The minimum output entropy conjecture
and its stronger (majorization) version are then stated and explained. In
Sec.~\ref{s:gen} we analyze the two channel maps in detail, and develop
some useful properties of their output entropies. In
  Sec.~\ref{s:gauss} we prove the minimum output entropy conjecture for
the restricted scenario in which only Gaussian states may be fed into the
channel. In Sec.~\ref{s:lb} we present a collection of lower bounds on
$\ms$. These lower bounds are consistent with the minimum
output entropy conjecture.  Moreover, in the low and high noise regimes they
approach asymptotically the upper bounds from which the
conjecture arises. In Sec.~\ref{s:local} we obtain necessary conditions
on any input state that minimizes the output
entropy.  We demonstrate, in particular, that every coherent-state input
produces an output state which achieves a local minimum of the output
entropy.  Finally, in Sec.~\ref{s:maggiorizz} we address the stronger
version of the conjecture by exhibiting some evidence that output states
produced by coherent-state inputs majorize all other output
states.  The paper is structured so that
Sec.~\ref{s:gauss}--Sec.~\ref{s:maggiorizz} may be read independently. 
The most technical parts of the derivations have been relegated to
appendices.
\section{Channel models}\labell{s:nm}
We will consider two bosonic channels---the thermal-noise channel and
the classical-noise channel---both of which belong to the class of
Gaussian CP-maps, i.e, they evolve Gaussian inputs states into Gaussian
output states~\cite{werner}.  Although we will limit our attention to
single-mode channels, our results can be extended to the multimode case
(see, e.g.,~\cite{futuro}).  

\paragraph*{Thermal-noise channel:} 
Here the signal photons interact with an environment that is in
thermal equilibrium.   The channel can be considered to be a beam
splitter that couples the input state and the thermal reservoir to an
output port, with the input-state transmissivity of this beam splitter
being the channel's quantum efficiency {\mbox{$\eta\in[0,1]$}}. 
The CP-map  
${\cal E}_\eta^N$ for the thermal-noise channel is easily obtained by
tracing away the thermal state of the environment mode (annihilation
operator $b$, with average photon number
$N$) from the unitary evolution
\begin{eqnarray}
U=\exp\left[(ba^\dag-b^\dag a)\arctan\sqrt{\frac{1-\eta}\eta}\:\right]\;
\labell{defu}.  
\end{eqnarray}
This unitary evolution leads to the beam splitter transformation
\begin{eqnarray}\left\{\begin{array}{l}
a\longrightarrow U^\dag aU=\sqrt{\eta}\; a + \sqrt{1-\eta} \; b\cr\cr
b\longrightarrow U^\dag bU=\sqrt{\eta}\; b - \sqrt{1-\eta} \; a,
\end{array}\right.
\labell{uno}
\end{eqnarray}
where $a$ is the annihilation operator for the channel mode. The
thermal-noise channel's CP-map is thus
\begin{eqnarray}
\ce_\eta^N(\rho)=\mbox{Tr}_b\left[U\:\rho\otimes\tau_b\:U^\dag\right]
\;\labell{mappa}
\end{eqnarray}
where $\rho$ is the channel input state (of the mode $a$) and
$\tau_b=[N/(N+1)]^{b^\dag b}/(N+1)$ is the environment thermal state.
The case $N=0$ (zero-temperature reservoir) represents the pure-loss
channel, in which each input photon has probability $\eta$ of reaching the
output. At positive temperatures the noise source is active, 
injecting noise photons into the channel mode.  For $\eta=1$, the CP-map
is the identity: the reservoir does not couple to the output, hence the
$\eta = 1$ channel is noiseless.

\paragraph*{Classical-noise channel:} Here a classical Gaussian
noise is superimposed on the transmitted field, i.e., the classical-noise
channel is characterized by the CP-map~\cite{hall}
\begin{eqnarray}
{\cal N}_n (\rho)=\int {\rm d}^2 \mu \;
P_{n}(\mu) \; D(\mu)\rho D^{\dagger}(\mu)
\labell{due}
\end{eqnarray} 
where
\begin{eqnarray}
P_n(\mu)=\frac{e^{-|\mu|^2/n}}{\pi n},
\labell{tre}
\end{eqnarray}
and $D(\mu)\equiv \exp(\mu a^\dag - \mu^* a)$ is the displacement
operator for the mode $a$.  The classical-noise channel's
CP-map~(\ref{due}) is unital---it maps the identity into
the identity---whereas that for the $\eta <1$ thermal-noise channel is
not.  Nevertheless, the classical-noise channel can be seen to be a
limiting case of the thermal-noise channel in which the field
transformation~(\ref{uno}) is replaced by
\begin{eqnarray}
a\longrightarrow a+\mu
\;\labell{campo2},
\end{eqnarray}
where $\mu$ is a classical, complex-valued random variable
distributed according to Eq.~(\ref{tre}). The map $\cn_n$ is then
obtained from $\ce_\eta^N$ in the limit $\eta\to 1$ and
$N\to\infty$, with $(1-\eta)N\to n$ [see Eq.~(\ref{characteristic}),
below].  For $n=0$ the CP-map~(\ref{due}) becomes the identity, while for
$n>0$ it injects noise photons into the channel.  

A more realistic
construct for the classical-noise channel is the following.  We first
propagate the input state through a thermal-noise channel with transmissivity $\eta$
and $N=0$.  Then---to compensate for the propagation loss---we employ a phase-insensitive
amplifier of gain $\kappa = 1/\eta$ \cite{caves}.  For $\eta = 1/(n+1)$ the concatenation of these two maps exactly yields the classical-noise channel ${\cal N}_n$ (see Eq.~(\ref{Bnew})).

In response to a vacuum-state input, $|0\rangle$, both the thermal-noise
and the classical-noise channels produce thermal output states
given by the density operator
\begin{eqnarray}
\rho'_0\equiv \frac{1}{M+1}\left(\frac{M}{ M+1}
\right)^{a^\dag a}
\labell{vacuum}\;,
\end{eqnarray} 
with $M=(1-\eta)N$ for the thermal-noise channel ${\cal E}_{\eta}^N$, and
$M=n$ for the classical-noise channel ${\cal N}_n$.  The von Neumann
entropy of
$\rho'_0$ is easily found to be $S(\rho'_0)\equiv
-$Tr$[\rho'_0\ln\rho_0']=g(M)$, with~\cite{werner,yuen}
\begin{eqnarray}
g(x)\equiv(1+x) \ln(1+x) -x \ln x
\labell{gfun}
\end{eqnarray}
for $x>0$ and $g(0)=0$ (see App.~\ref{s:vacuum}).  The output state that
results from the coherent-state input $|\alpha\rangle$ can be obtained by
the following displacement-operator transformation of 
$\rho'_0$,
\begin{eqnarray}
\rho'_{\alpha}= D(\alpha')\; \rho'_0 \; D^{\dag}(\alpha'),
\labell{coher}
\end{eqnarray}
where $\alpha'=\sqrt{\eta}\:\alpha$ for the thermal-noise channel and
$\alpha'=\alpha$ for the classical-noise channel [see Eqs.~(\ref{cov1})
and (\ref{cov}), below].  Clearly, the states $\rho'_{\alpha}$
and $\rho'_0$ have identical von Neumann entropies as the former is a
unitary transformation of the latter~\cite{chuang}.

\subsection{Minimum output entropy}
We are interested in the minimum von Neumann
output entropy $\ms$, which is defined to be
\begin{eqnarray}
{\mathbb S}({\cal M}) \equiv \min_{\rho \in {\cal H}}  \;S({\cal
    M}(\rho))  \;,
\labell{def}
\end{eqnarray}
where $\cm$ is the channel's CP-map and $\cal H$ is the Hilbert space of
the information carrier.  The concavity of the von Neumann
entropy~\cite{chuang,wehrl} implies that $\ms$ in
Eq.~(\ref{def}) can be achieved with a pure-state input.  Moreover, given
two CP-maps, $\cm_1$ and $\cm_2$, we have that
\begin{eqnarray}
\ms(\cm_2\circ\cm_1)\geqslant \ms(\cm_2)
\;\labell{min},
\end{eqnarray}
where the composition
$(\cm_2\circ\cm_1)(\rho)\equiv\cm_2(\cm_1(\rho))$ is the map in which
$\cm_2$ acts on the output of $\cm_1$. Inequality~(\ref{min}) can be shown
by noting that every possible input to $\cm_2$ on the left-hand side is included in the
minimization that is implicit on the right-hand side.
\paragraph*{{\bf Conjecture (i):}}
The minimum output entropies for the thermal-noise and
classical-noise channels are achieved by coherent-state inputs, so
that
\begin{eqnarray}
{\mathbb S}= 
\left\{
\begin{array}{cccc}
g((1-\eta)N)& & \mbox{for }{\cal E}_{\eta}^{N}& \\ \\ 
g(n) & & \mbox{for }\cn_n&.\\ 
\end{array}
\right. 
\labell{minent}
\end{eqnarray}
{\em Discussion}:  Because the two entropies on the right-hand side of (\ref{minent}) are
achieved by coherent-state inputs, they immediately provide
upper bounds for the minimum output entropies of the thermal-noise and
classical-noise channels, respectively.  Thus the conjecture states that
they are also lower bounds.  The conjecture is trivially satisfied by the
zero-temperature ($N=0$) thermal-noise channel,  because
the purity of a coherent state is preserved under the action of the
loss map~\cite{c}.  The conjecture is again trivially satisfied by the
$\eta=0$ thermal-noise channel, because the map
$\ce_0^N$ sends every input state into a thermal state with average photon
number $N$. 

A physical justification for our conjecture resides in the fact that
in each channel the input state is contaminated by noise from a reservoir
whose quantum phase is completely random. [The reservoirs are the
thermal state
$\tau_b$ for $\ce_\eta^N$ and the
classical random source associated with the distribution $P_n(\mu)$ for
$\cn_n$]. One thus expects  that the extraction of any coherence
from the reservoir---which could be used to reduce the output entropy below
the level when no photons are sent through the channel---will be
impossible.  Some preliminary results in this sense were obtained
in~\cite{paz}, where it was shown that, for the thermal-noise map, the
linearized entropy is minimized by a vacuum-state input in the limit of
low coupling $\eta\ll 1$ and high temperature $N\gg 1$. In the sections
to come we will provide further evidence in support of this conjecture.   We will also investigate 
the following stronger version of Conjecture~(i). 

\paragraph*{{\bf Conjecture (ii):}}  The output
states produced by coherent-state inputs majorize all other
output states. \\ 
{\em Discussion}:   By definition, a state $\rho$ majorizes a state
$\sigma$ (a property which we denote by $\rho\succ\sigma$) if all
ordered sums of the eigenvalues of $\rho$ equal or exceed the
corresponding sums for
$\sigma$~\cite{chuang}, i.e.,
\begin{eqnarray}
\rho\succ\sigma\Longleftrightarrow\sum_{i=0}^q\lambda_i\geqslant\sum_{i=0}^q\mu_i
\quad\forall q\ge 0
\;\labell{maj},
\end{eqnarray}
where $\lambda_i$ and $\mu_i$ are the eigenvalues of $\rho$ and
$\sigma$, respectively, arranged in decreasing order (e.g.,
$\lambda_0\geqslant\lambda_1\geqslant\cdots$). If $\rho\succ\sigma$
then $S(\rho)\leqslant S(\sigma)$, so that Conjecture~(ii) implies
Conjecture~(i). The converse is generally not true, in that 
$S(\rho)\leqslant S(\sigma)$ does not guarantee that
{\mbox{$\rho\succ\sigma$}}. A necessary and sufficient condition for a
state $\rho$ to majorize a state $\sigma$ is that $\sigma$ 
be obtainable from $\rho$ by the action of a unital map~\cite{chuang}.
Conjecture~(ii) for the thermal-noise channel would then be proved if, for
each input state
$\rho$, we could find a unital map $\cal L_\rho$ such that
$\ce_\eta^N(\rho)={\cal
L}_\rho(\rho_0')$, where $\rho'_0$ from Eq.~(\ref{vacuum}) is the output
state that is due to a vacuum-state input. For the classical-noise
channel proof we would want a unital map satisfying 
$\cn_n(\rho)={\cal L}_\rho(\rho_0')$.  We postpone further discussion of 
Conjecture~(ii) until Sec.~\ref{s:maggiorizz}, where we will present some
evidence that supports its validity.  The next four sections will
concentrate on Conjecture~(i).

\section{Channel properties}\labell{s:gen}
In this section we develop some useful properties of the thermal-noise
and classical-noise bosonic channels, and we provide some
insights into their output entropies.

\subsection{Covariance}\labell{s:cov}
From (\ref{due}), it is easy to
show that the classical-noise channel's CP-map is covariant under
displacement, i.e.,
\begin{eqnarray}
\cn_n(D(\alpha)\:\rho\:D^\dag(\alpha))=
D(\alpha)\:\cn_n(\rho)\:D^\dag(\alpha)
\;\labell{cov1}.
\end{eqnarray}
The thermal-noise channel's CP-map enjoys a similar relation,
viz.,
\begin{eqnarray}
{\cal E}_\eta^N(D(\alpha)\rho D^\dag(\alpha))&=&
D(\sqrt{\eta}\:\alpha)\:{\cal E}_\eta^N(\rho)\:
D^\dag(\sqrt{\eta}\:\alpha)\labell{cov},
\end{eqnarray}
where the $\sqrt{\eta}$ factor comes from the beam-splitter
transformation~(\ref{uno}). Moreover, the circular symmetries of the
probability distribution $P_n$ of Eq.~(\ref{tre}) and of the environment thermal
state $\tau_b$ imply that both noise maps are invariant under the
action of a phase shift transformation $e^{i\phi a^\dag a}$.  From
these two properties plus the unitary invariance of the von
Neumann entropy~\cite{wehrl}, it follows that any two input states
that differ by a displacement and/or a phase-shift transformation
will produce output states with the same entropy. In particular, this
means that all coherent states produce the same output entropy, as
discussed previously.

\begin{figure}[hbt]
\begin{center}
\epsfxsize=1.
\hsize\leavevmode\epsffile{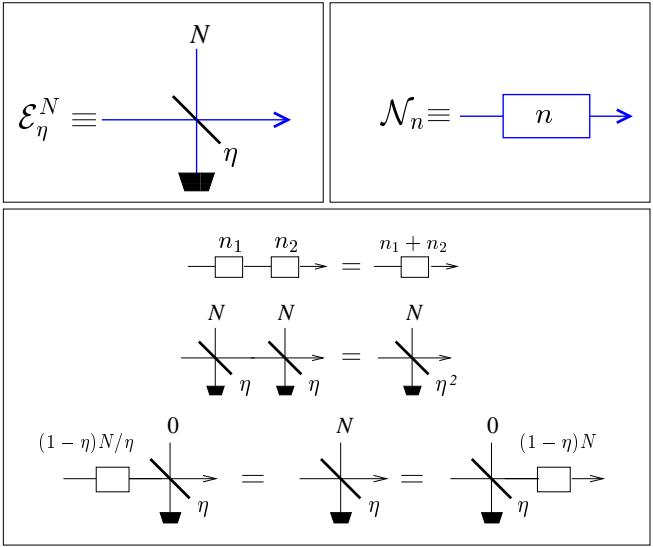}
\end{center}
\caption{Top panel:~graphical representations of the thermal-noise and
classical-noise channel's CP-maps; 
  the signal photons propagate from left to right in these diagrams.
Bottom panel:~composition
  rules for these channel models; from top to bottom are representations of
  Eqs.~(\ref{pr1}), (\ref{pr2}), and~(\ref{pr3}), respectively.}
\labell{f:composition}\end{figure}
\subsection{Composition rules}\labell{s:composition}
A complete description of the channels ${\cal E}_\eta^N$ and   ${\cal
N}_n$ is provided by the transformation of the symmetrically ordered
characteristic function of the input state~\cite{walls}
\begin{eqnarray}
\chi({\mu})\equiv
\mbox{Tr}[\rho D(\mu)]\;\labell{chi},
\end{eqnarray}
from which $\rho$ is recovered as $\int {\rm d}^2\mu\:\chi(\mu)\:D(-\mu)/\pi$.
Evaluated on the $\rho'$, state at the channel output, this function becomes
\begin{eqnarray}
\chi'(\mu)=
\left\{
\begin{array}{llll}
\chi\left(\sqrt{\eta}\mu\right)\:e^{-(1-\eta)(N+1/2)|\mu|^2}
& &\mbox{for}&{\cal E}_\eta^N\\\\
\chi\left(\mu\right)\:e^{-n|\mu|^2}& &\mbox{for}&{\cal N}_n\;.
\end{array}\right.\labell{characteristic}\;
\end{eqnarray}
As a consequence of the Gaussian character of this evolution, thermal
states (for which $\chi(\mu)$ is proportional to a zero-mean Gaussian
distribution) evolve into thermal states.  The following composition
rules, summarized in Fig.~\ref{f:composition}, follow immediately from
Eq.~(\ref{characteristic}):
\begin{eqnarray}
{\cal N}_{n_2}\circ{\cal N}_{n_1}&=&{\cal N}_{n_1+n_2}\;,\labell{pr1}\\
{\cal E}_{\eta_2}^{N_2}\circ{\cal E}_{\eta_1}^{N_1}&=&
{\cal
  E}_{\eta_1\eta_2}^{N'}\;,\ 
\;\labell{pr2}
\end{eqnarray}
where $N'=[{\eta_2(1-\eta_1)N_1+(1-\eta_2)N_2}]/({1-\eta_1\eta_2})$.
From Eq.~(\ref{pr1}) we see that concatenating two
identical classical-noise channels yields another classical-noise
channel that is twice as random.  From Eq.~(\ref{pr2}) we see that
concatenating two identical thermal-noise maps results in another
thermal-noise channel with the same reservoir, but whose
transmissivity has been squared.  It is also possible, using
Eq.~(\ref{characteristic}), to express the thermal-noise
channel as either a pure-loss channel followed by a classical-noise
channel, or vice versa:  
\begin{eqnarray}
{\cal E}_\eta^N&=&{\cal N}_{(1-\eta)N}\circ{\cal E}_\eta^0=
{\cal E}_\eta^0\circ{\cal N}_{(1-\eta)N/\eta}\;\labell{pr3}.
\end{eqnarray}
From these equations,
some useful properties of the output entropy can be derived. In
particular:\begin{itemize}
\item Because a unital channel increases entropy (the output always
  majorizes the input~\cite{chuang}), Eq.~(\ref{pr1}) 
  implies that the entropy $S({\cal N}_n(\rho))$ at the output of the
  classical channel is an increasing function of $n$, i.e., for any
  $\rho$ and $\Delta\geqslant 0$  
\begin{eqnarray}
S(\cn_{n+\Delta}(\rho))\geqslant S(\cn_n(\rho))
\;\labell{re1}.
\end{eqnarray}
\item Because ${\cal N}_n$ is unital, we can also infer that the
  entropy $S({\cal E}_\eta^N(\rho))$ at the output of the thermal-noise
  channel is an increasing function of $N$ . This follows because for any
$\rho$ and
$\Delta\geqslant 0$  we have that
\begin{eqnarray}
\lefteqn{S({\cal E}_\eta^{N+\Delta}(\rho))= }\nonumber \\
& & S(({\cal N}_{\Delta(1-\eta)}\circ{\cal E}_\eta^{N})(\rho))\geqslant
S({\cal E}_\eta^{N}(\rho))
\;\labell{dim},
\end{eqnarray}
where Eq.~(\ref{pr1}) and the first equality of Eq.~(\ref{pr3}) have
been used.
\item Using (\ref{pr2}) with $N_1=N_2=N$ in conjunction with
  relation~(\ref{min}) shows that the minimum output entropy of the
thermal-noise 
  channel is a decreasing function of $\eta$, 
\begin{eqnarray}
\ms(\ce_{\eta}^N)\geqslant\ms(\ce_{\eta'}^N)\quad{\mbox{for }}\eta'\geqslant\eta
\;\labell{rel}.
\end{eqnarray}
Note, however, that the output entropy $S(\ce_\eta^N(\rho))$ is {\em
  not} a decreasing function of $\eta$ for every $\rho$. This is
because the thermal-noise map does necessarily increase
the entropy of the input. Consider what happens when the channel input is 
a thermal state with average photon number $N_0$ satisfying
$N_0>N$, so that $g(N_0)>g(N)$ holds. According to
Eq.~(\ref{characteristic}), the output state is a thermal state
with average photon number $\eta N_0+(1-\eta)N<N_0$.  Its
entropy is therefore $g(\eta N_0+(1-\eta)N)$, which is smaller than
$g(N_0)$ and is an increasing function of $\eta$.
\item A stronger version of (\ref{rel})  can be obtained by using
  relation (\ref{pr2}) with $N_1 \neq N_2$. In this case,
 (\ref{min}) implies
\begin{eqnarray}
\ms(\ce_{\eta}^{N})\geqslant\ms(\ce_{\eta'}^{N'})
\quad{\mbox{for  }}\eta'\geqslant\eta\;,
\;\labell{rel3}
\end{eqnarray}
and $N\geqslant\frac{1-\eta'}{1-\eta}N'$.
\item The transmissivity inequality in (\ref{rel}) can be inverted if the
thermal
  photon numbers are appropriately chosen, viz.,
\begin{eqnarray}
\ms(\ce_\eta^N)&\geqslant&\ms(\ce_{\eta'}^{N'})\quad{\mbox{for }}\eta\geqslant\eta'
\;\labell{qqq},\end{eqnarray}
where now $N'\leqslant\frac{(1-\eta)N+\eta'-\eta}{1-\eta'}$. This relation
is proven in App.~\ref{s:ampli} and, together with (\ref{dim}),
(\ref{rel}), and (\ref{rel3}), is illustrated in Fig.~\ref{f:region}.
\item Using the first equality in Eq.~(\ref{pr3}), along with
  (\ref{min}), we can establish the following relation between the minimum
output entropies of the classical-noise map and the thermal-noise map,
\begin{eqnarray}
{\mathbb S}({\cal E}_{\eta}^{N})= {\mathbb S}({\cal N}_{(1-\eta)N}\circ{\cal
  E}_\eta^0)\geqslant {\mathbb S}({\cal N}_{(1-\eta)N})\;
 \labell{ins1}.
\end{eqnarray}
Physically, this says that deleting the
pure-loss beam-splitter map ${\cal E}^0_\eta$ can only decrease the
output noise.  An important consequence of (\ref{ins1}) is
that if Conjecture~(i) holds for the classical-noise channel, then it
must also hold for the thermal-noise channel.
\item The reverse
counterpart of (\ref{ins1}) is given by
\begin{eqnarray}
{\mathbb S}({\cal N}_n)\geqslant{\mathbb S}({\cal E}^{(n-n')/n'}_{1-n'})
\;\labell{ins2},
\end{eqnarray}
for all $n'\in[0,\min\{1,n\}]$ (see App.~\ref{s:ampli} for the
derivation).  The key consequence of (\ref{ins2}) is that if 
Conjecture~(i) is true for the thermal-noise channel, then it is must
also be valid for the classical-noise channel.
\end{itemize}

\begin{figure}[hbt]
\begin{center}
\epsfxsize=.8
\hsize\leavevmode\epsffile{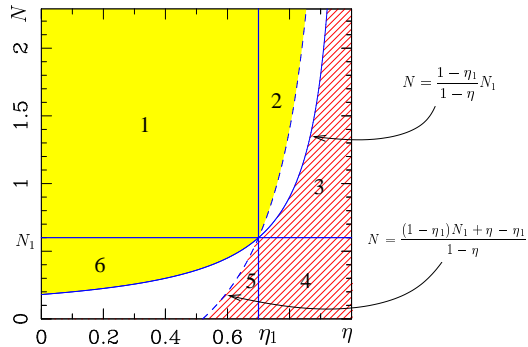}
\end{center}
\caption{Plot of the minimum entropy regions for the thermal-noise channel
  as constrained by (\ref{dim})--(\ref{qqq}).
  Each $(\eta,N)$ point corresponds to a different
  thermal-noise channel described by the CP-map $\ce_\eta^N$.  Given a
channel
  with transmissivity $\eta_1$ and average thermal photon-number $N_1$,
the gray
  (hatched) region represents channels whose minimum entropies are greater
  (less) than $\ms(\ce_{\eta_1}^{N_1})$. [The line $(\eta,0)$ has
  minimum entropy zero, and belongs to the hatched region]. In the
  white regions, the composition rules from this section do not
establish the relation between $\ms(\ce_\eta^N)$ and 
$\ms(\ce_{\eta_1}^{N_1})$; in
  Fig.~\ref{f:regione1} these regions will be partially filled by
  exploiting the lower bounds that will be introduced in
  Sec.~\ref{s:lb}. Showing that the
  upper white region is gray and the lower white region is hatched would complete
the proof of Conjecture~(i).  Regions 1 and
  4 follow from (\ref{dim}) and~(\ref{rel}), regions 3 and 6
  follow from (\ref{rel3}), while regions 2 and 5 are consequences of 
  (\ref{qqq}).  The plot assumes $\eta_1=0.7$ and $N_1=0.6$. }
\labell{f:region}\end{figure}
\section{Gaussian input states}\labell{s:gauss}
Here we show that Conjecture~(i) is true if
we restrict the channel input to be a Gaussian state $\rho_G$. Such
an input has a symmetric characteristic function that is a Gaussian
form~\cite{werner},
\begin{eqnarray}
\chi(\mu)&=&\exp\left[-\zeta_0\cdot \zeta^\dag-\frac 12
  \zeta\cdot\Gamma\cdot \zeta^\dag\right] 
\qquad \zeta\equiv(\mu^*,-\mu)
\nonumber\labell{gaussia},
\end{eqnarray}
which is fully characterized by its first moment
$\zeta_0\equiv(\langle a\rangle,\langle a^\dag\rangle)$ and its
covariance matrix
\begin{eqnarray}
\Gamma\equiv\left[\begin{array}{cc}
\langle\{\Delta a,\Delta a^\dag\}\rangle/2&
\langle(\Delta a)^2\rangle\cr
\langle(\Delta a)^2\rangle&
\langle\{\Delta a,\Delta a^\dag\}\rangle/2
\end{array}\right]
\;\labell{correlation},
\end{eqnarray}
(here $\langle\;\cdot\;\rangle\equiv \mbox{ Tr}[\;\cdot\;\rho_G]$ is 
expectation with respect to $\rho_G$, $\Delta a \equiv a-\langle
a\rangle$, and
$\{\;\cdot\;,\;\cdot\;\}$ denotes the anticommutator).  The coherent
state $|\alpha\rangle$ is Gaussian with
$\Gamma=\openone/2$ and $\zeta_0=(\alpha,\alpha^*)$. The entropy of
$\rho_G$ depends only on its covariance matrix~\cite{werner,sohma}
and is equal to $g(\sqrt{\det\Gamma}-1/2)$.  Both  $\ce_\eta^N$
and $\cn_n$ transform Gaussian input states into Gaussian output states.
Moreover, by means of Eq.~(\ref{characteristic}),
evolution under these CP-maps transforms covariance matrices according to 
\begin{eqnarray}
\Gamma\longrightarrow\Gamma'=\left\{\begin{array}{lll}
\Gamma+n\openone&&\mbox{for }\cn_n\cr\cr
\eta\Gamma+(1-\eta)(N+1/2)\openone&&\mbox{for }\ce_\eta^N
\;,\end{array}\right.
\;\labell{gammaprimo}
\end{eqnarray}
and first moments according to $\zeta_0\longrightarrow \zeta_0$ for
$\cn_n$, and
$\zeta_0\longrightarrow\sqrt{\eta}\zeta_0$ for $\ce_\eta^N$.  The output
entropy of a Gaussian input state is, hence, equal to
$g(\sqrt{\det\Gamma'}-1/2)$, which is always greater than or equal to
the output entropy of the vacuum, i.e.,  the right-hand-side of
Eq.~(\ref{minent}), as we now will show. 

For the classical map $\cn_n$ it is
possible to rewrite (\ref{gammaprimo}) as 
\begin{eqnarray}
\det\Gamma'&=&\det\Gamma+n\left(n+\langle\{\Delta a,\Delta
a^\dag\}\rangle\right)
\;\labell{gammacl},
\end{eqnarray}
which is always greater than $(n+1/2)^2$ because $\langle\{\Delta
a,\Delta a^\dag\}\rangle\geqslant 1$ and, from the strong version of
the uncertainty relation~\cite{holevo2,sakurai}, $\det\Gamma\geqslant
1/4$. In other words, we have that
\begin{eqnarray}
S(\cn_n(\rho_G))&=&g\left(\sqrt{\det\Gamma+n\left(n+\langle\{\Delta a,\Delta
a^\dag\}\rangle\right)}-1/2\right)\nonumber\\&\geqslant& g(n)
\;\labell{other}.
\end{eqnarray}

Likewise, we see that Conjecture~(i) is true for the thermal-noise channel
$\ce_\eta^N$ whose input is limited to be a Gaussian state because
\begin{eqnarray}
\det\Gamma'&=&\eta^2\det\Gamma+(1-\eta)(N+1/2)[(1-\eta)(N+1/2)
\nonumber\\&&+\eta\langle\{\Delta a,\Delta a^\dag\}\rangle]
\;\labell{gammath},
\end{eqnarray}
which implies $\det\Gamma'\geqslant[(1-\eta)N+1/2]^2$.

\section{Lower bounds}\labell{s:lb}
In this section we present some lower bounds on the output
entropy. These bounds are consistent with Conjecture~(i), and
collectively they are asymptotically tight in the limits of low and high
noise.  We will treat the two channel models in succession, starting with
the classical noise case. 

\begin{figure}[hbt]
\begin{center}
\epsfxsize=.8
\hsize\leavevmode\epsffile{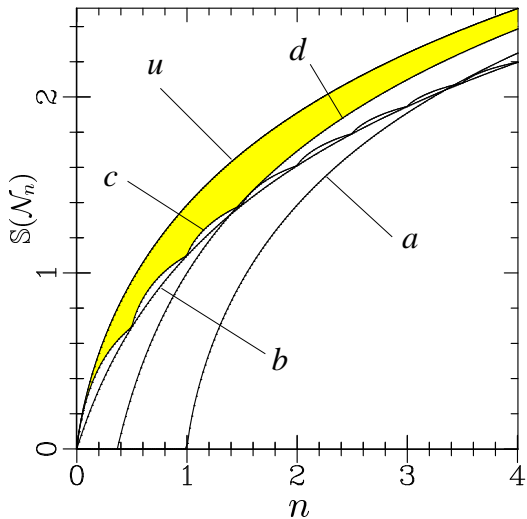}
\end{center}
\caption{Bounds on the minimum output entropy of the classical-noise
channel $\cn_n$ versus average photon number of the classical noise, $n$.
Curves {\it a, b, c, } and {\it d}
are the lower bounds given in (\ref{f1}), (\ref{f2}),
  (\ref{f3}), and (\ref{f4}), respectively. The upper bound $u$ is the
  function $g(n)$. The minimum entropy $\ms(\cn_n)$ is constrained by
these bounds to lie in
  the gray region, and is required to be an increasing function of $n$.
Conjecture~(i) states that $\ms(\cn_n)=g(n)$.}
\labell{f:bounds}\end{figure}

\subsection{Classical-noise channel}\labell{s:clas}
Because $g(n)$ is the output entropy that results when the input is a
coherent state, $g(n)$ is an upper bound on the minimum output
entropy $\ms({\cn_n})$ of the classical-noise channel. Four different
lower bounds on $\ms({\cn_n})$ are given below.  As seen in
Fig.~\ref{f:bounds}, bound~{\it a} is implied by bound~{\it d} and bound~{\it
b} is implied by  bound~{\it c}.  Nevertheless, we explain all of them
because the derivations of {\it a} and {\it b} are simpler.  In the
limits of low and high values of the noise parameter
$n$, it can be shown that this collection of bounds is asymptotically
tight, i.e., 
$\lim_{n\rightarrow 0} \ms_c(\cn_n)/g(n) = \lim_{n\rightarrow\infty}
\ms_d(\cn_n)/g(n) = 1$, where $\ms_j(\cn_n)$ denotes bound~{\it j}. 

\paragraph*{Lower bound a:} By considering the Husimi function of the
output state, we find that for $n\geqslant 1$
\begin{equation}
\ms(\cn_n)\geqslant g(n-1)
\;\labell{f1},
\end{equation}
by the following argument.
Any initial state $\rho$ can
be written as~\cite{walls}
\begin{eqnarray}
\rho=\int {\rm d}^2\alpha\; Q(\alpha)\:\sigma(\alpha)
\;\labell{husimi},
\end{eqnarray}
where $Q(\alpha)\equiv\langle\alpha|\rho|\alpha\rangle/\pi$ is that 
state's Husimi function and
\begin{eqnarray}
\sigma(\alpha)=\int \frac{{\rm d}^2\lambda}\pi
D(\lambda)\;e^{\lambda^*\alpha-\lambda\alpha^*+|\lambda|^2/2} 
\;\labell{husimi1}.
\end{eqnarray}
Under the action of the map $\cn_n$, the state $\rho$ evolves to
\begin{eqnarray}
\cn_n(\rho)=\int {\rm d}^2\alpha\; Q(\alpha)\:\cn_n(\sigma(\alpha))
\;\labell{husimi2}.
\end{eqnarray}
The operator $\cn_n(\sigma(\alpha))$ is not in general a quantum
state. However, for $n>1$ it is a displaced thermal state with
average photon number $n-1$, i.e.,
\begin{eqnarray}
\cn_n(\sigma(\alpha))&=&D(\alpha)\int \frac{{\rm d}^2\mu}\pi
\:\frac{e^{-\frac{|\mu|^2}{n-1}}}{n-1}|\mu\rangle\langle\mu|
\;D^\dag(\alpha)\nonumber\\
&=&D(\alpha)\frac 1n\left(\frac {n-1}n\right)^{a^\dag a}D^\dag(\alpha)
\;\labell{husimi3},
\end{eqnarray} 
which has entropy $g(n-1)$. Lower bound~{\it a} then can be obtained from
Eq.~(\ref{husimi2}) because
$Q(\alpha)$ is a probability distribution, and the von Neumann
entropy is concave~\cite{wehrl}.

\paragraph*{Lower bound b:} By considering the R\'enyi
entropy $S_2(\rho')\equiv -\ln$Tr$[(\rho')^2]$ calculated on the
channel output $\rho'=\cn_n(\rho)$, we find
\begin{equation}
\ms(\cn_n)\geqslant\ln(2n+1)
\;\labell{f2},
\end{equation}
via two simple steps.
As discussed in~\cite{renyinostro}, it is possible to show that
$S_2(\rho')$ achieves its minimum \mbox{$\ln(2n+1)$} when the channel
input is a coherent state. Lower bound~(\ref{f2}) is then a trivial
consequence of the von Neumann entropy $S$ being greater than
or equal to the R\'enyi entropy $S_2$,~\cite{zyc}.

\paragraph*{Lower bound c:} By using a more sophisticated connection
between the von Neumann entropy and the R\'enyi entropy~\cite{renyi,wang},
we find for $k\geqslant 1$ an integer and \mbox{$n\in[(k-1)/2, k/2]$}
\begin{equation}
\ms(\cn_n)\geqslant
-\lambda_k(n)\ln\lambda_k(n)
-[1-\lambda_k(n)]\ln\frac{1-\lambda_k(n)}k
\;\labell{f3},
\end{equation}
where
\begin{eqnarray}
\lambda_k(n)=\frac{1-\sqrt{1-(k+1)(1-k/(2n+1))}}{k+1}
\;\labell{lambda}.
\end{eqnarray}
The derivation of (\ref{f3}) is postponed to
App.~\ref{s:app1}.

\paragraph*{Lower bound d:} Using the properties of the map $\cn_n$ we
find
\begin{equation}
\ms(\cn_n)\geqslant 1+\ln n
\;\labell{f4},
\end{equation}
as we now demonstrate.
Consider a pure state $|\psi\rangle$ which generates an output state with
spectral decomposition 
\begin{eqnarray}
\rho'=\cn_n\left(|\psi\rangle\langle\psi|\right)=\sum_k\gamma_k|
\gamma_k\rangle\langle\gamma_k|\;\labell{image},
\end{eqnarray}
where $\{\gamma_k\}$ is a probability distribution and
$\{|\gamma_k\rangle\}$ are the orthonormal eigenvectors. From
the definition~(\ref{due}) of the classical-noise channel's CP-map, we
have
\begin{eqnarray}
\gamma_k=\langle\gamma_k|\rho'|\gamma_k\rangle=
\int {\rm d}^2\mu\;P_n(\mu)
\Big|\langle\gamma_k|D(\mu)|\psi\rangle\Big|^2
\;\labell{asa}.
\end{eqnarray}
The quantity $\left|\langle\gamma_k|D(\mu)|\psi\rangle\right|^2$ is a
probability distribution over $k$, and
$\left|\langle\gamma_k|D(\mu)|\psi\rangle\right|^2/\pi$ is a probability
distribution over $\mu$~\cite{nota}.  Therefore,
the convexity of $x^z$ for $z\ge 1$ ensures that
\begin{eqnarray}
\lefteqn{\mbox{Tr}[(\rho')^{z}]=\sum_k\left(\int {\rm d}^2\mu\;P_n(\mu)
\left|\langle\gamma_k|D(\mu)|\psi\rangle\right|^2\right)^{z}
\;}
\nonumber \\ &\leqslant&\sum_k\int \frac{{\rm d}^2\mu}\pi\;\Big(\pi
  P_n(\mu)\Big)^{z} 
\left|\langle\gamma_k|D(\mu)|\psi\rangle\right|^2.\labell{ottusolor}
\end{eqnarray}
Because $P_n(\mu)$ is a Gaussian, it follows that
\begin{eqnarray}
\Big(\pi P_n(\mu)\Big)^{z}=\frac{\pi
P_{n/{z}}(\mu)}{{z}\: n^{{z}-1}}\;\labell{boh},
\end{eqnarray}
and the right-hand-side of (\ref{ottusolor}) can be rewritten in
terms of the image of $|\psi\rangle$ under the action of the map
$\cn_{n/{z}}$, i.e.,
\begin{eqnarray}
\mbox{Tr}[(\rho')^{z}]\leqslant\frac
{\mbox{Tr}[\cn_{n/{z}}(|\psi\rangle\langle\psi|)]}
{{z}\:n^{{z}-1}}=\frac 1{{z}\: n^{{z}-1}}
\;\labell{boh1}.
\end{eqnarray}
This relation can be used to calculate a lower bound for the von
Neumann entropy by observing that~\cite{zyc}
\begin{eqnarray}
S(\rho')&=&\lim_{{z}\to
  1}-\frac{\ln\mbox{Tr}[(\rho')^{z}]}{{z}-1}
\geqslant\lim_{{z}\to 1}\frac{\ln({z}\:
  n^{{z}-1})}{{z}-1}\nonumber\\
&=&
1+\ln n
\;\labell{ecco}.
\end{eqnarray}
Inequality~(\ref{ecco}) applies for any pure state $|\psi\rangle$, so we
conclude that (\ref{f4}) holds.

\begin{figure}[hbt]
\begin{center}
\epsfxsize=.8
\hsize\leavevmode\epsffile{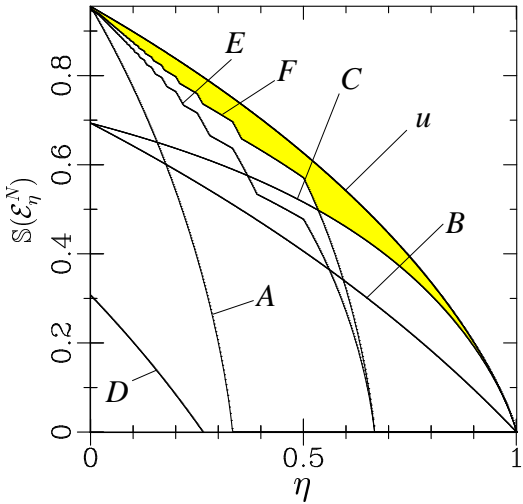}
\end{center}
\caption{Bounds on the minimum output entropy of the thermal-noise channel
$\ce_\eta^N$ as functions of channel transmissivity $\eta$ for $N=1/2$.
Curves {\it A}, {\it E}
  and {\it F} are the lower bounds (\ref{A}), (\ref{E}),
  and~(\ref{F}), respectively [Here {\it E} is the maximum over $k$ of
  the right-hand side of (\ref{E}) and {\it F} is the
maximum over
  $k$ of the right-hand side of (\ref{F})]. Curves {\it B}, {\it
    C}, and {\it D} are the lower bounds (\ref{f2}), (\ref{f3}), and
(\ref{f4}), respectively, with
$n=(1-\eta)N$. The
  upper bound $u$ is the function $g((1-\eta)N)$.  The minimum
output entropy
  $\ms(\ce_\eta^N)$ is constrained to lie in the gray region, and is
required to be a
  decreasing function of $\eta$. Conjecture~(i) states that
  $\ms(\ce_\eta^N)=g((1-\eta)N)$.}  \labell{f:boundsth}\end{figure}

\subsection{Thermal-noise channel}
The same techniques that we used to derive lower bounds for the
classical-noise channel can also be employed for the thermal-noise channel
$\ce_\eta^N$.  The bounds we obtain in this case are reported in
Figs.~\ref{f:boundsth}, \ref{f:boundsth1}, and~\ref{f:regione1}.
\paragraph*{Lower bound A:} 
Repeating the Husimi function calculation employed above, we find 
\begin{equation}
\ms(\ce_\eta^N)\geqslant g((1-\eta)N-\eta)
\;\labell{A},
\end{equation}
as follows.  
We replace Eq.~(\ref{husimi3}) with
\begin{eqnarray}
\lefteqn{\ce_\eta^N(\sigma(\alpha))=}\;\\\nonumber
&&D(\sqrt{\eta}\:\alpha)\int \frac{{\rm d}^2\mu}\pi
\:\frac{e^{-\frac{|\mu|^2}{(1-\eta)N-\eta}}}{(1-\eta)N-\eta}|\mu\rangle\langle\mu|
\;D^\dag(\sqrt{\eta}\:\alpha)
,\labell{husimi3p}
\end{eqnarray}
which for $(1-\eta)N\geqslant\eta$ is a displaced thermal state whose
entropy equals $g((1-\eta)N-\eta)$. Bound~{\it A} then follows from
the concavity of the von Neumann entropy.
\paragraph*{Lower bounds B, C, and D:} A simple strategy to
derive bounds on $\ms(\ce_\eta^N)$ is to exploit 
relation~(\ref{ins1}), which links the minimum output entropy of
$\ce_\eta^N$ to that of $\cn_{(1-\eta)N}$.  Thus, by replacing
$n$ with
$(1-\eta)N$ inequalities~(\ref{f2}), (\ref{f3}) and (\ref{f4}) of
bounds {\it b, c, } and {\it d } immediately become bounds {\it B},
{\it
  C}, and {\it D} on $\ms(\ce_\eta^N)$, respectively, which we have
plotted versus channel transmissivity $\eta$ in Fig.~\ref{f:boundsth}.
[Applying this same method to lower bound {\it a}, we obtain
$g((1-\eta)N-1)$, which is not useful, as it is already implied by
(\ref{A}).]
\paragraph*{Lower bound E:}  A further lower bound for
the thermal-noise channel can be derived from the
properties of the beam splitter; for all integer $k$ it states that
\begin{equation}
\ms(\ce_\eta^N)\geqslant \left\{
\begin{array}{ll}\frac {k-1}k\:g\left(\frac k{k-1}(1-\eta)
N\right)&\mbox{for }\eta\leqslant\frac 1k
\\\\
\frac{k-1}k\:g\left(\frac
  k{k-1}\left[(1-\eta)N-\eta+\frac 1k\right]\right)
&\mbox{for
}\eta\geqslant\frac 1k\:,
\end{array}\right.
\;\labell{E}
\end{equation}
the proof appears in App.~\ref{s:bs}. Curve {\it E} in
Fig.~\ref{f:boundsth} is the maximum over $k$ of the right-hand side of
(\ref{E}).
\paragraph*{Lower bound F:} A more sophisticated version of bound
{\it E} is given by
\begin{equation}
\ms(\ce_\eta^N)\geqslant \left\{
\begin{array}{ll}\frac {k-1}k\:g\left((1-\eta)N\right)
+\frac 1k \ms(\cn_{(1-\eta)N})&\mbox{for }\eta\leqslant\frac 1k
\\\\
\frac{k-1}k\:g\left((1-\eta)N-\eta+\frac 1k\right)
&\\&\mbox{for
}\eta\geqslant\frac 1k\:,\\
\quad\qquad+\frac 1k \ms(\cn_{(1-\eta)N-\eta+1/k})
&
\end{array}\right.
\;\labell{F}
\end{equation}
which is also proven in App.~\ref{s:bs}. Even though 
$\ms(\cn_n)$ is not known, (\ref{F}) provides a usable lower bound
for $\ms(\ce_\eta^N)$ when we replace $\ms(\cn_n)$ with any of the lower
bounds discussed in Sec.~\ref{s:clas}. Curve {\it F} in
Fig.~\ref{f:boundsth} is the maximum
over $k$ of the right-hand side of (\ref{F}).  If
Conjecture~(i) is true, then (\ref{F}) becomes an equality for
$\eta\leqslant 1/k$ , i.e., this bound is tight. The same is not true for
the bound~(\ref{E}).

\begin{figure}[hbt]
\begin{center}
\epsfxsize=1.
\hsize\leavevmode\epsffile{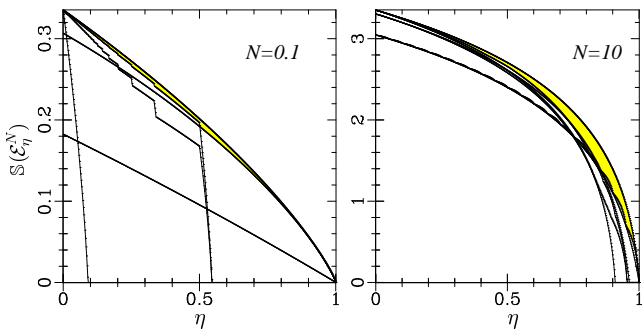}
\end{center}
\caption{Same as Fig.~\ref{f:boundsth} but for different values of the
  parameter $N$: in the left plot $N=0.1$, in the right plot $N=10$.
  At both high and low average thermal photon numbers,
  the greatest of these lower bounds approaches the upper bound.}
\labell{f:boundsth1}\end{figure}

\begin{figure}[hbt]
\begin{center}
\epsfxsize=.7
\hsize\leavevmode\epsffile{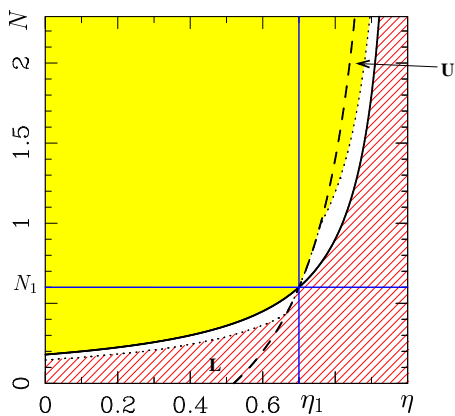}
\end{center}
\caption{Plot of the minimum entropy regions  for the thermal-noise
channel
  for the same parameters used in Fig.~\ref{f:region}. Here, the region
  {\bf L} has been added to the hatched region of Fig.~\ref{f:region}
  by comparing the lower bounds {\it A}--{\it F} of the channel
  $\ce_{\eta_1}^{N_1}$ with the upper bound $g((1-\eta)N)$ for generic
  $\eta$ and $N$: if the former is larger than the latter, we can
  conclude that $\ms(\ce_{\eta_1}^{N_1})$ is greater than
  $\ms(\ce_{\eta}^N)$.  Likewise, the region {\bf U} has been added
  to the gray region of Fig.~\ref{f:region}, by comparing the upper
  bound on $\ms(\ce_{\eta_1}^{N_1})$ with the lower bounds on
  $\ms_(\ce_\eta^N)$.} \labell{f:regione1}\end{figure}

\section{Necessary conditions for minimum output entropy}\labell{s:local}
In this section we study the conditions that an input state must to
satisfy in order to provide a (local or global) minimum for the output
entropy.  We will show that coherent-state inputs satisfy all of these
conditions for both the classical-noise channel and the thermal-noise
channel.  We begin, in the first subsection, by deriving
a local minimum condition from the directional
derivative of the output entropy.  Then, in the second subsection, we 
generalize this result into a more stringent condition for minimality.  

\subsection{Local minimum condition}
Given a CP-map $\cm$, a necessary condition for an input state
$\sigma_0$ to provide a local minimum of the output entropy
$S(\cm(\rho))$ can be obtained from the directional
derivatives of this quantity~\cite{werner}.  Given $t\in[0,1]$
and a generic state $\sigma$, this condition requires that 
\begin{eqnarray}
&&\left.\frac{\partial}{\partial
    t}S(\cm(\sigma(t)))\right|_{t=0}\labell{minim}\\&&\qquad\qquad=
\mbox{Tr}\Big\{[\cm(\sigma_0)-\cm(\sigma)]\ln\cm(\sigma_0)\Big\}\geqslant
0
\;\nonumber,
\end{eqnarray}
where $\sigma(t)$ is the mixed state $(1-t)\:\sigma_0+t\sigma$. For
both the thermal-noise map $\ce_\eta^N$ and the classical-noise
map $\cn_n$, this condition is satisfied by the arbitrary coherent state
$\sigma_0=|\alpha\rangle\langle\alpha|$, as can be shown by using
covariance properties of the noise (see Sec.~\ref{s:cov}) to rewrite the
output entropy derivative as follows:
\begin{eqnarray}
&&\left.\frac{\partial}{\partial
    t}S(\cm(\sigma(t)))\right|_{t=0^+}=
\mbox{Tr}[(\rho_0'-\cm(\tilde\sigma))
\ln\rho_0']\nonumber\\&&
\qquad=\mbox{Tr}
[a^\dag a(\cm(\tilde\sigma)-\rho_0')]\ln\frac M{M+1}
\nonumber\\&&\qquad=\zeta
\mbox{Tr}[a^\dag a\:\tilde\sigma]\ln\frac M{M+1}\geqslant
0
\;\labell{minim1},
\end{eqnarray}
where $\zeta=1$ for $\cn_n$ and $\zeta=\eta$ for $\ce_\eta^N$,
$\rho_0'$ is the output state 
Eq.~(\ref{vacuum}) generated by using a vacuum state input, and
$\tilde\sigma$ is the state
$D^\dag(\alpha)\:\sigma\:D(\alpha)$. The last equality in
(\ref{minim1}) derives from
\begin{eqnarray}
\mbox{Tr}\left[a^\dag a\:\cm(\tilde\sigma)\right]=
\zeta\mbox{Tr}[a^\dag a\:\tilde\sigma]+M
\;\labell{minim2},
\end{eqnarray}
which holds because $M$ is the average number of photons in the
state~$\rho_0'$. Physically, the inequality in (\ref{minim1}) is a
consequence of the vacuum's being the input state that produces the output
state with lowest average photon number.
\subsection{Eigenvalue minimum condition}
In deriving (\ref{minim}) we required that the entropy be
locally increasing when moving along the trajectory $\sigma(t)$,
whose intermediate states are all mixed.  A more stringent requirement
follows from using the pure-state trajectory
$\sigma_\theta=|\sigma_\theta\rangle\langle\sigma_\theta|$ with
\begin{eqnarray}
|\sigma_\theta\rangle=\cos\theta\:|\sigma_0\rangle+\sin\theta\:|\sigma_\perp\rangle
\;\labell{traiett},
\end{eqnarray}
where $\sigma_0\equiv|\sigma_0\rangle\langle\sigma_0|$ is the
putative minimizing state for the output entropy, and
$|\sigma_\perp\rangle$ is any state that is orthogonal to
$|\sigma_0\rangle$. Expanding the output state entropy in a Taylor's
series, we have
\begin{eqnarray}
\lefteqn{S(\cm(\sigma_\theta))=}
\nonumber \\ && S(\cm(\sigma_0))+\theta
\left.\frac\partial{\partial\theta'}S(\cm(\sigma_{\theta'}))\right|_{\theta'=0}+\cdots
\;\labell{derivative}.
\end{eqnarray}
[The entropies of the channels we are considering are differentiable.]
If the state $|\sigma_0\rangle$ is a local minimum for the output
entropy, then the term that is linear in $\theta$ must vanish. This
requirement implies the following necessary condition for local minimality
\begin{eqnarray}
\lefteqn{
\left.\frac\partial{\partial\theta}S(\cm(\sigma_\theta))\right|_{\theta=0}=
} \nonumber \\ &&
-2\mbox{Re}\langle\sigma_\perp|\cm^*\left(\ln\cm(\sigma_0)\right)|\sigma_0\rangle=0
\;\labell{con},
\end{eqnarray}
where $\cm^*$ is the dual map associated with $\cm$, such that for any
two operators $A$ and $B$,
Tr$\left[A\cm(B)\right]=$Tr$\left[\cm^*(A)B\right]$. Inasmuch as
this condition must be valid for all $|\sigma_\perp\rangle$, we can
conclude that the operator 
\begin{eqnarray}
{\cal F}({\sigma_0})\equiv-\cm^*(\ln\cm(\sigma_0))\;\labell{caleffe}
\end{eqnarray}
has $|\sigma_0\rangle$ as an eigenvector. The properties of ${\cal
  F}(\sigma_0)$ guarantee that the eigenvalue associated with 
$|\sigma_0\rangle$ is its output entropy.  Shor and Ruskai~\cite{shorf}
have a different way of introducing the operator ${\cal F}(\sigma_0)$ to
study the minima of a channel's output entropy; their approach 
does not require the output entropy to be differentiable.

For the channels we are considering, the vacuum input
evolves into the thermal state $\rho_0'$ of Eq.~(\ref{vacuum}), so that
\begin{eqnarray}
{\cal F}(|0\rangle\langle 0|)=\cm^*(a^\dag a\:\ln\frac{M+1}M+\ln (M+1)) 
\;\labell{cf1}.
\end{eqnarray}
In particular, for the classical-noise channel this quantity simplifies
appreciably because $\cn_n$ is its own dual. Indeed, because
$\cn_n(a^\dag a)=a^\dag a+n$, we find
\begin{eqnarray}
{\cal F}(|0\rangle\langle 0|)=
g(n)\openone+a^\dag a\;\ln \frac {n+1}n
\;\labell{cf},
\end{eqnarray}
which shows that the vacuum is an eigenvector of   ${\cal
F}(|0\rangle\langle 0|)$, and hence satisfies the  local minimum
condition of Eq.~(\ref{con}).  The positivity of
the operator $a^\dag a$ implies that the vacuum is also the
eigenvector with minimum eigenvalue. Using the definition of $\cal F$,
Eq.~(\ref{cf}) can be used to express the output entropy of a general
pure-state input $\rho$ as
\begin{eqnarray}
S(\cn_n(\rho))=g(n)+E\ln\frac{n+1}n-S(\cn_n(\rho)\:\|\:\rho_0')
\;\labell{cf3},
\end{eqnarray}
where $E=$Tr$[a^\dag a\rho]$ is the average photon number of the
input~$\rho$ and
$S(\rho_1\|\rho_2)\equiv$Tr$[\rho_1(\ln\rho_1-\ln\rho_2)]$ is the
relative entropy between states $\rho_1$ and
$\rho_2$~\cite{chuang}. This equation allows us to restate 
Conjecture~(i) in the form
\begin{eqnarray}
E\ln\frac{n+1}n\geqslant S(\cn_n(\rho)\:\|\:\rho_0')
\;\labell{conjecture2}.
\end{eqnarray}
Proving this relation for all $\rho$ is equivalent to proving 
Conjecture~(i).  For coherent-state inputs, we can use the
covariance properties of the noise under displacements, to show that
\begin{eqnarray}
{\cal F}(|\alpha\rangle\langle\alpha|)=
D(\alpha){\cal F}(|0\rangle\langle 0|)D^\dag(\alpha)
\;\labell{rr},
\end{eqnarray}
which guarantees that the coherent state $|\alpha\rangle$ is an
eigenvector of~${\cal F}(|\alpha\rangle\langle\alpha|)$. If we could
prove that the coherent states are the only ones which satisfy this
condition, we would have succeeded in proving the conjecture: the
coherent states would be the only states that satisfy the necessary
condition~(\ref{con}) for minimality.  Unfortunately, such is not the
case because Fock states $|n\rangle$ are also eigenvectors of the
corresponding ${\cal F}(|n\rangle\langle n|)$; this follows from the
states $\cn_n(|n\rangle\langle n|)$ being diagonal in the Fock
basis~\cite{caves1}. Fock states other than the vacuum are not, however,
minima for the output entropy, as discussed in Sec.~\ref{s:maggiorizz}.
Note that condition~(\ref{caleffe}) was first introduced in
Ref.~\cite{hall}, where it was claimed that Conjecture~(i) was proven.
A more careful analysis of \cite{hall} reveals a
fundamental missing link in that proof:  even though it is
shown that number states satisfy condition~(\ref{caleffe})
and that the vacuum is the number state with the lowest output entropy, it is
not proven that they are the {\it only} states that satisfy this condition.
Hence, there is still a possibility that another state (with output
entropy lower than the vacuum) might satisfy the condition.

The case of the thermal-noise channel can be treated in a similar manner,
showing that here too coherent states are eigenvectors of the
corresponding $\cal F$ operators. In this case, the dual map of
$\ce_\eta^N$ can be written as
\begin{eqnarray}
(\ce_\eta^N)^*(\rho)=\mbox{Tr}_b[(\openone_a\otimes\tau_b)U^\dag
(\rho\otimes\openone_b) U]
\;\labell{duale},
\end{eqnarray}
which is unital and satisfies
\begin{eqnarray}
(\ce_\eta^N)^*(a^\dag a)=\eta a^\dag a+(1-\eta)N
\;\labell{duale1}.
\end{eqnarray}
Using these properties and Eq.~(\ref{cf1}), we find
\begin{eqnarray}
{\cal F}(|0\rangle\langle 0|)=
g((1-\eta)N)\openone+\eta a^\dag a\;\ln \frac {(1-\eta)N+1}{(1-\eta)N}
\;\labell{cf12}.
\end{eqnarray}
As in the case of Eq.~(\ref{cf}), the vacuum is the eigenvector with
minimum eigenvalue of the operator $\cal F$. Moreover, 
Eq.~(\ref{rr}) applies here, i.e., the coherent state $|\alpha\rangle$
is an eigenvector of ${\cal F}(|\alpha\rangle\langle\alpha|)$, and
Eq.~(\ref{cf3}) becomes
\begin{eqnarray}
S(\ce_\eta^N(\rho))&=&g((1-\eta)N)+\eta 
E\ln\frac{(1-\eta)N+1}{(1-\eta)N}\nonumber\\
&&-S(\ce_\eta^N(\rho)\:\|\:\rho_0')
\;\labell{cf31}.
\end{eqnarray}

\section{Majorization}\labell{s:maggiorizz}
In the previous sections we gave some justifications in support of 
Conjecture~(i). Here we focus on the stronger version of this
conjecture, i.e., Conjecture~(ii).  We begin by presenting evidence that 
the output states generated by coherent states
majorize the ones generated by Fock states. 
\subsection{Fock-state inputs}
Because the output states generated by the
coherent state are all unitarily equivalent, we can focus on the
ordered eigenvalue sums for the thermal state $\rho_0'$ from
Eq.~(\ref{vacuum}), i.e.,
\begin{eqnarray}
\frac 1{M+1}\sum_{i=0}^q\left(\frac M{M+1}\right)^i=1-\left(\frac
  M{M+1}\right)^{q+1}
\;\labell{somma}.
\end{eqnarray}
The ordered partial sums~(\ref{somma}) for all $q$ must be compared with 
their Fock-state-input counterparts. In the
case of classical noise, these can be numerically evaluated by
observing that, for a Fock input state $|k\rangle$, the output is
diagonal in the Fock basis and takes the form~\cite{caves1}
\begin{eqnarray}
\cn_n(|k\rangle\langle k|)&=&\sum_{i=0}^\infty
\lambda_{i}\:|i\rangle\langle i|
\;\labell{caves},
\end{eqnarray} 
where
\begin{eqnarray}
\lambda_{i}\equiv
\sum_{j=0}^{\min(k,i)}\left(\begin{matrix}i\\
      j\end{matrix}\right)\left(\begin{matrix}k\\j
\end{matrix}\right)
\frac {n^{k+i-2j}}{(n+1)^{k+i+1}}
\;.\;\labell{jacobi}
\end{eqnarray}
[Note that in the case $k=0$, Eq.~(\ref{caves}) reduces to the
vacuum evolution Eq.~(\ref{vacuum})].  Evaluation of the
ordered partial sums is tedious but can be performed analytically. In
particular, for $k=1$ and $n\geqslant 1$, the ordered
partial sums $\{\Sigma_q\}$ of the first $q+1$ eigenvalues are
\begin{eqnarray}
\Sigma_q=1-\left(1+\frac{q+1}{n(n+1)}\right)\left(\frac{n}{n+1}\right)^{q+1}
\;\labell{sigmaq},
\end{eqnarray}
which, for all $q$, are smaller than the corresponding sums in Eq.~(\ref{somma}). The case $n<1$
is analogous: for sufficiently large values of $q$ the sum is the same as in Eq.~(\ref{sigmaq}),
while for small $q$ it is given by
\begin{eqnarray}
\Sigma_q=1-\left(1+\frac{q+2}{n(n+1)}\right)\left(\frac{n}{n+1}\right)^{q+2}-\frac
n{(n+1)^2}
\labell{sigmaq1},
\end{eqnarray}
which again is smaller than the sum in Eq.~(\ref{somma}).
\begin{figure}[hbt]
\begin{center}
\epsfxsize=1.
\hsize\leavevmode\epsffile{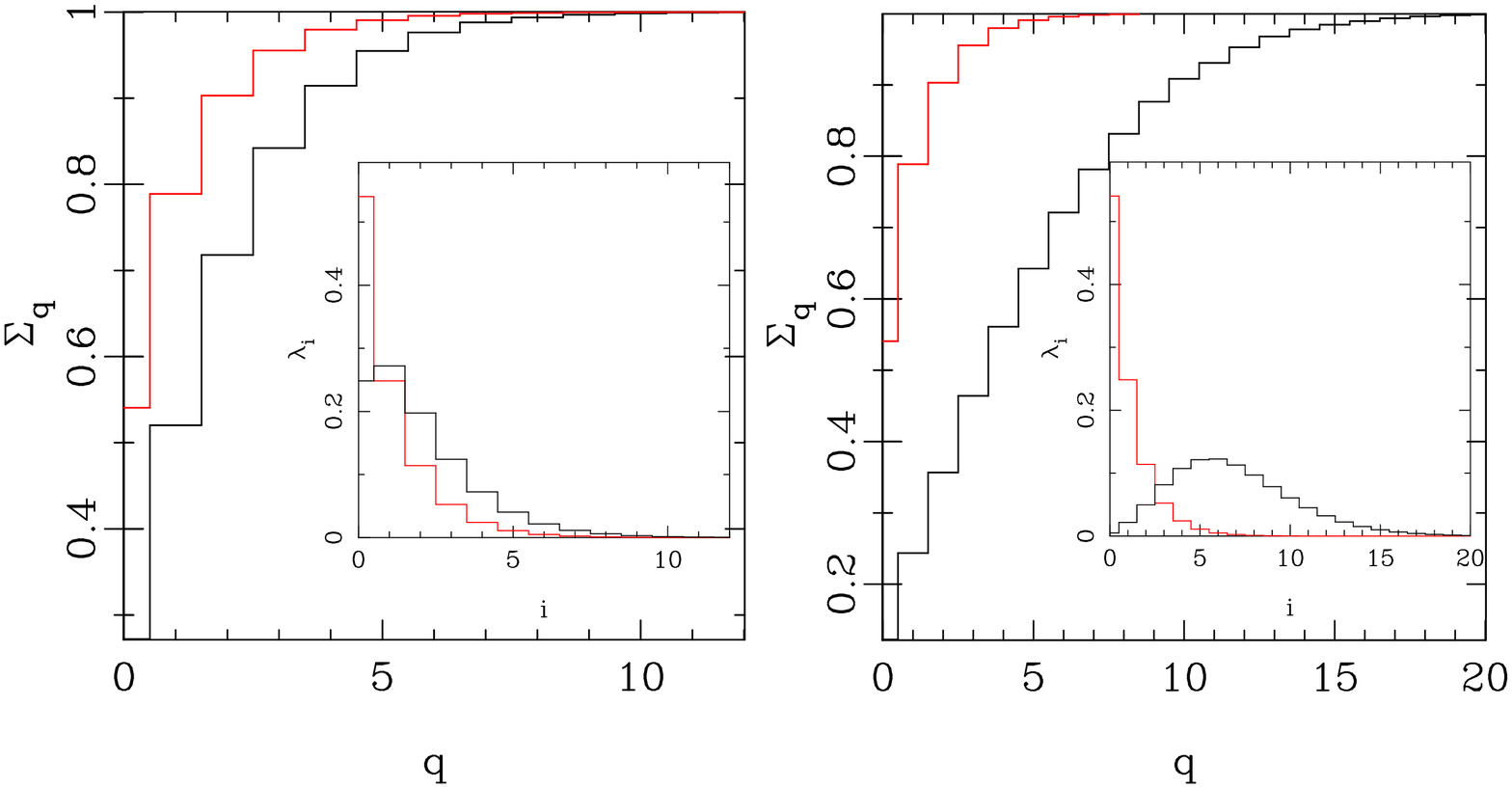}
\end{center}
\caption{Majorization analysis of the map
  $\cn_n$. Plots of $\Sigma_q$, the sum of the $q+1$ largest
  eigenvalues, for the thermal state $\rho_0'$ (gray staircase) and the evolved Fock state
$\cn_n(|k\rangle\langle k|)$ (black
  staircase). In both cases the thermal state majorizes the evolved
Fock state, which
implies that the entropy of the former is
  smaller than the entropy of the latter~\cite{chuang}. The insets show
  the eigenvalues $\{\lambda_i\}$ as functions of the photon
  number $i$.  Here $k=1$, $n=0.85$ for the left plot, and $k=6$, $n=0.85$
  for the right plot.}  \labell{f:maggiorizz}\end{figure} 
In
Fig.~\ref{f:maggiorizz} a numerical comparison between
Eq.~(\ref{somma}) and the ordered partial sums of the eigenvalues from
Eq.~(\ref{caves}) is presented for two specific cases; $\rho_0'$ majorizes
$\cn_n(|k\rangle\langle k|)$ in both.

The same analysis can be repeated in the case of the thermal channel,
observing that
\begin{eqnarray}
\ce_\eta^0(|k\rangle\langle k|)=\sum_{m=0}^kp_{m}\:|m\rangle\langle
m|,
\end{eqnarray}
where $\{p_m\}$ is the binomial distribution
\begin{eqnarray}
p_{m}=\left(\begin{matrix}k\\
    m\end{matrix}\right) {\eta^m(1-\eta)^{k-m}}
\;.
\end{eqnarray}
Using Eq.~(\ref{caves}) and the decomposition~(\ref{pr3}) the thermal
evolution of the Fock state $|k\rangle$ can be calculated as
\begin{eqnarray}
\ce_\eta^N(|k\rangle\langle k|)=
\sum_{m=0}^kp_{m}\;
\cn_{(1-\eta)N}(|m\rangle\langle m|)
\;.
\nonumber\\\labell{formulazza}
\end{eqnarray}
Note that the output is again diagonal in the Fock basis. Moreover,
if the output of the vacuum majorizes the output of the other Fock
states for the classical channel $\cn_n$, Eq.~(\ref{formulazza}) can
be used to prove that this must also be true for the thermal channel
$\ce_\eta^N$. In fact, if $\cn_n(|0\rangle\langle 0|)$ majorizes
$\cn_n(|m\rangle\langle m|)$, then there exists~\cite{chuang} a unital
map ${\cal L}_m$ such that $\cn_n(|m\rangle\langle m|)={\cal
  L}_m\left(\cn_n(|0\rangle\langle 0|)\right)$. Now, because
$\ce_\eta^N(|0\rangle\langle 0|)=\cn_{(1-\eta)N}(|0\rangle\langle
0|)$, Eq.~(\ref{formulazza}) implies that
\begin{eqnarray}
\ce_\eta^N(|k\rangle\langle k|)=
\sum_{m=0}^kp_{m}\;
{\cal L}_m\left(\ce_\eta^N(|0\rangle\langle 0|)\right)
\;.
\nonumber\\\labell{formulazza1}
\end{eqnarray}
The convex sum of unital maps is a unital map, hence
$\ce_\eta^N(|k\rangle\langle k|)$ is obtained from a unital
transformation of $\ce_\eta^N(|0\rangle\langle 0|)$, which
implies~\cite{chuang} that the latter majorizes the former.

\subsection{Arbitrary input states}
Further insight into Conjecture~(ii) is provided by generalizing our analysis to the case of an arbitrary pure-state input.  Because it is sufficient to establish Conjecture~(ii) for the classical-noise channel, we shall only consider that case.  As shown in App.~\ref{s:numbrep}, see also \cite{caves1}, when an arbitrary pure state $|\psi\rangle$ is fed into the classical-noise map $\cn_n$, the resulting output state $\rho'$ has the following Fock-state representation,
\begin{eqnarray}
\lefteqn{\langle k+l|\rho'|k\rangle = \sqrt{\frac{k!}{(k+l)!}}\sum_{j=0}^\infty
\sqrt{\frac{j!}{(j+l)!}}\,\psi_{j+l}\psi_j^*} \nonumber \\
&\times&\frac{(j+k+l)!}{j!k!}\frac{n^{j+k}}{(1+n)^{j+k+l+1}} \nonumber \\ 
&\times& F(-j,-k;-(j+k+l);1-n^{-2}),
\labell{eq:numbrep}
\end{eqnarray}
for $k,l\ge 0$, where $\{\psi_n\}$ are the Fock-state coefficients of $|\psi\rangle$, and $F(\alpha,\beta;\gamma;z)$ is the hypergeometric function.

To probe the output-state eigenvalue behavior associated with an
arbitrary pure-state input with up to 10 photons, we used the
following procedure.  Eleven complex numbers, $\{\,\phi_n : 0\le n\le
10\,\}$, whose real and imaginary parts were randomly distributed on
the interval [-1, 1] were used to generate a pure state, via
\begin{equation}
|\psi\rangle = \sum_{n=0}^{10}\phi_n |n\rangle/\sqrt{\sum_{n=0}^{10}|\phi_n|^2}\;.
\end{equation}
Using the $\{\psi_n\}$ for this state we diagonalized $\rho'$---found
from Eq.~(\ref{eq:numbrep}) truncated to the Fock states
$\{\,|n\rangle : 0\le n\le 40\,\}$---and then calculated the ordered
eigenvalue sum, $\Sigma_q$, for $0\le q \le 40$.  In all 100 trials of
this procedure, we found that the output state generated by the vacuum
majorized the output state produced by the arbitrary pure state.
Figure~\ref{f:purestatemaj} shows this comparison for five of our 100
trials.  These particular input states had average amplitudes $1.805
-0.002i$, $1.318 - 0.340i$, $1.596 + 0.404i$, $1.810 - 0.255i$, $1.546
+ 0.276i$, and average photon numbers 5.059, 3.375, 4.976, 4.748,
4.163, respectively.
\begin{figure}[hbt]
\begin{center}
\vspace*{.5cm}
\epsfxsize=.6
\hsize\leavevmode\epsffile{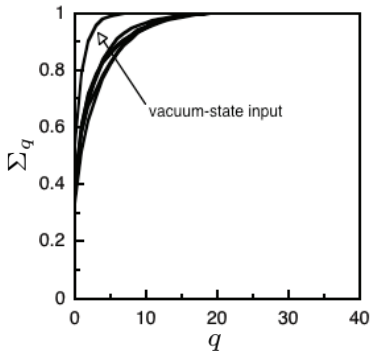}
\end{center}
\vspace*{-.5cm}
\caption{Majorization analysis of the map
 $\cn_n$ for $n=0.85$. Plots of the sum of the $q+1$ largest
  eigenvalues for the thermal state $\rho_0' = \cn_n(|0\rangle \langle 0|)$ and the evolved pure state
$\cn_n(|\psi\rangle\langle \psi|)$ for five randomly-generated $|\psi\rangle$. In all cases the thermal state majorizes the evolved
$|\psi\rangle$ state, which
implies that the entropy of the former is
  smaller than the entropy of the latter~\cite{chuang}.} \labell{f:purestatemaj}
\end{figure} 

\subsection{Simulated annealing optimization}
As a final test of Conjecture~(ii), we used simulated
annealing~\cite{simanneal}---a well known technique for finding global
extrema---to minimize the classical-noise channel's output entropy.
As in the previous subsection, the input state was truncated to lie in
$\mbox{span}\{\,|n\rangle : 0 \le n \le 10\,\}$, the output states
were constrained to lie in $\mbox{span}\{|n\rangle : 0 \le n \le
40\,\}$, and we limited our consideration to the classical-noise
channel.  A variety of initial pure-state inputs were employed, in
conjunction with an exponential cooling schedule.  In all cases, the
resulting minimum output entropy was extremely close to that achieved
by a vacuum-state input.  Indeed, in every case the associated input
state---at the end of the simulated annealing iterations---was very
nearly a coherent state.  Figure~\ref{f:simanneal1} shows the
progression of output entropy values for the $n=0.85$ classical-noise
channel when the simulated annealing procedure was initiated with the
Fock-state input $|6\rangle$ and 400 iterations were performed.  The
initial output entropy in this run was 3.754; the final output entropy
in this run was 1.846.  The latter is very close to $g(0.85) = 1.841$,
which is the output entropy for a coherent-state input.  The final
input state $|\psi\rangle_{400}$, after the 400 iterations, had mean
amplitude $-0.116 + 1.861i$, average photon number 3.47, and 99.9\%
overlap with the coherent state $|\alpha\rangle$ for $\alpha = -0.12 +
1.88i$, viz., $|\langle\alpha|\psi\rangle_{400}|^2 = 0.999$.  For this
Fock-state input, we found that the output-state eigenvalues at every
iteration majorized those for preceding iterations, see, e.g.,
Fig.~\ref{f:simanneal1}.
\begin{figure}[hbt]
\begin{center}
\vspace*{.5cm}
\epsfxsize=1
\hsize\leavevmode\epsffile{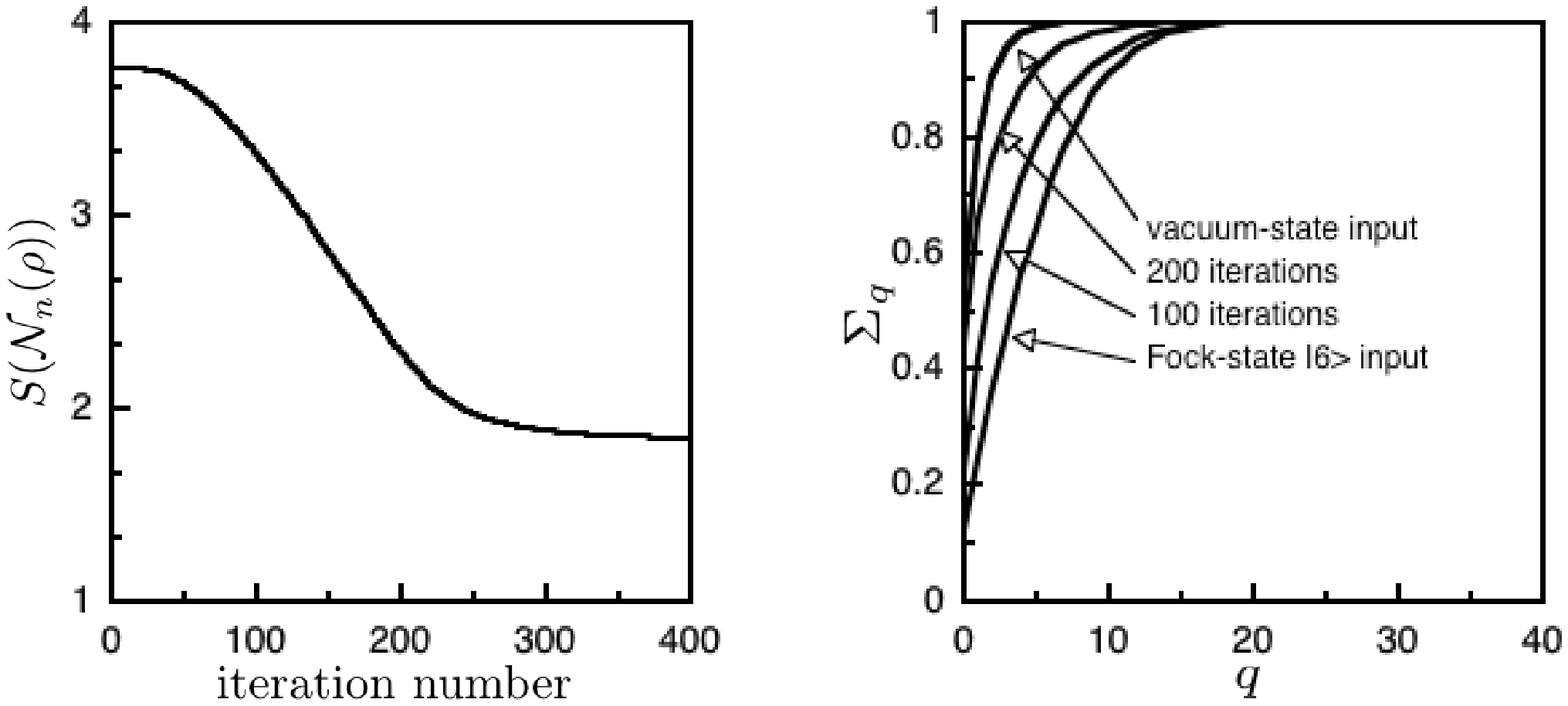}
\end{center}
\vspace*{-.5cm}
\caption{Left:  simulated annealing minimization of the classical-noise channel's output entropy for 
$n=0.85$.  Plot of the output entropy $S({\cal N}_n(\rho))$ versus iteration number when the minimization algorithm is initiated with the Fock-state input $|6\rangle$. Right:  majorization analysis of the map
 $\cn_n$ for $n=0.85$.  Plots of $\Sigma_q$ (the sum of the $q+1$ largest
  eigenvalues) for the thermal state $\rho_0' = \cn_n(|0\rangle \langle 0|)$, the evolved Fock state $|6\rangle$, and the output states obtained after 100, 200, and 400 iterations of the simulated annealing algorithm.  The curves for the vacuum-state input and the output state after 400 iterations are indistinguishable on this scale.} \labell{f:simanneal1}
\end{figure} 

\section{Conclusions}\labell{s:concl}
We conjectured that the minimum entropies at the output of
two Gaussian bosonic channels (with thermal or classical
noise) are achieved by inputs that are coherent states.  Physically, this
conjecture is plausible: the complete absence of correlation between the
input state and the channel's environment state would seem to forbid the
existence of an input whose injection reduces the output entropy to a
level lower than that achieved when no photons are transmitted. In support of
our conjecture, we presented four separate arguments.  First, we proved
that the conjecture is true when we restrict the analysis to
Gaussian-state inputs. Second, we established a suite of lower
bounds on the minimum output entropy, which are all  
compatible with the conjecture and which show that the conjecture is
asymptotically correct at low and high noise levels.  Third, we studied
local minimum conditions on the output entropy; input coherent states
were shown to be local minima that satisfy the operator identities
which are necessary conditions for minimality.  Fourth, we analyzed a
stronger version of the conjecture, namely that the output state produced
by a coherent-state input majorizes all other output states. In
support of this stronger conjecture, we presented evidence for
number-state inputs and randomly-selected inputs. In a companion 
paper~\cite{renyinostro}, we show that coherent states minimize the
output R\'enyi and Wehrl entropies for the classical-noise and
thermal-noise channels, thus lending further credence to the conjecture
in the present work.

\appendix
\section{Vacuum output entropy}\labell{s:vacuum}
Here we derive the output entropies of the channels $\cn_n$
and $\ce_\eta^N$ for vacuum-state inputs. Both channels evolve the vacuum
into the thermal state $\rho'_0$ of Eq.~(\ref{vacuum}).  For the
classical-noise channel, this can be seen by expressing $\rho_0'$ in terms
of coherent states, i.e.
\begin{eqnarray}
\cn_n(|0\rangle\langle 0|)&=& \int {\rm d}^2 \mu \;
P_{n}(\mu) \; |\mu\rangle \langle \mu| \\
&=&\frac 1{n+1}\left(\frac n{n+1}\right)^{a^\dag a}.
\;\labell{vacuum1}
\end{eqnarray}
The same relation applies for the thermal-noise channel, as can be seen by
using the decomposition~(\ref{pr3}) and the fact that the pure-loss
channel maps the vacuum state into itself, viz.,
\begin{eqnarray}
&&\ce_\eta^N(|0\rangle\langle 0|)=
\cn_{(1-\eta)N}\left[\ce_\eta^0(|0\rangle\langle 0|)
\right]=
\cn_{(1-\eta)N}(|0\rangle\langle 0|)\;.\nonumber\\&&
\labell{vacuum2}
\end{eqnarray}
The entropy of $\rho_0'$ is
\begin{eqnarray}
S(\rho'_0)&=&-\mbox{Tr}\left[
\rho'_0\left(a^\dag a\ln\frac M{M+1}-\ln(M+1)\right)\right]
\nonumber\\&=&(M+1)\ln(M+1)-M\ln M
\;\labell{va1},
\end{eqnarray}
where we have used the fact that the average photon number of $\rho'_0$ is
$M$.

\section{Proofs of (\ref{qqq}) and~(\ref{ins2})}\labell{s:ampli}
In order to prove (\ref{qqq}) and (\ref{ins2}), we employ the
amplifying channel defined by the field transformation
\begin{eqnarray}
a\longrightarrow\sqrt{\kappa}\:a+\sqrt{\kappa-1}\:c^\dag
\;\labell{amplification},
\end{eqnarray}
where $\kappa\geqslant 1$ is the amplifier gain and $c$ is the
annihilation operator of the amplifier's spontaneous-emission mode.  With
$c$ in the vacuum state, the symmetrically-ordered characteristic
function of the amplifying-channel map
${\cal
  A}_\kappa$ is easily shown to be 
\begin{eqnarray}
\chi'(\mu)=\chi(\sqrt{\kappa}\mu)\:e^{-(\kappa-1)|\mu|^2/2}
\;\labell{ampli}.
\end{eqnarray}
Linking Eqs.~(\ref{characteristic}) and (\ref{ampli}), we find the
decomposition rules
\begin{eqnarray}
\ce_\eta^N&=&\ce_{\eta'}^{N'}\circ{\cal A}_{\eta/\eta'}
\quad\quad\quad \mbox{for
}\eta\geqslant \eta'\;
\labell{composition5},\\
{\cal N}_n&=&{\cal E}_{1-n}^0\circ{\cal A}_{1/(1-n)}\quad\: \mbox{for
}n\leqslant 1
\;\labell{composition2},\\ 
{\cal N}_n &=& {\cal A}_{1/\eta} \circ {\cal E}^0_{\eta} \quad \mbox{for $n=(1-\eta)/\eta$}\;
\labell{Bnew},
\end{eqnarray}
where in~(\ref{composition5}),
\begin{eqnarray}
N'=\frac{(1-\eta)N+\eta'-\eta}{1-\eta'}\leqslant N\;.
\labell{qq}\;
\end{eqnarray}
Bound~(\ref{qqq}) now follows from combining
Eq.~(\ref{composition5}) with (\ref{min}), which establishes
that minimum entropy is increased by concatenation of two maps.  Using
relation~(\ref{composition2}) together with Eqs.~(\ref{pr1}) and
(\ref{pr3}), we obtain the identity
\begin{eqnarray}
{\cal N}_n&=&{\cal N}_{n-n'}\circ{\cal E}_{1-n'}^0\circ{\cal
  A}_{1/(1-n')}\nonumber\\&=&{\cal E}_{1-n'}^{(n-n')/n'}\circ{\cal
  A}_{1/(1-n')}
\;\labell{composition3},
\end{eqnarray}
which applies for $n'\in[0,\min(1,n)]$. Bound~(\ref{ins2}) follows
by removing the amplifier map using inequality~(\ref{min}).

\section{Derivation of lower bound~(\ref{f3})}\labell{s:app1}
In this appendix we derive lower bound~(\ref{f3}) for the minimum output
entropy $\ms(\cn_n)$. This bound arises from the connection between
the von Neumann entropy and the R\'enyi entropy of order two. Consider
the family of states $\rho$ with Tr$(\rho^2)=t$. As discussed
in~\cite{renyi}, the minimum values of $S(\rho)$ on this family are
obtained from states that have a non-degenerate eigenvalue $\lambda_0$
and a $k$-fold degenerate eigenvalue
$\lambda_1=(1-\lambda_0)/k\geqslant\lambda_0$, i.e.,
\begin{eqnarray}
S(\rho)&\geqslant&-\lambda_0\ln\lambda_0-(1-\lambda_0)\ln\frac{1-\lambda_0}k
\;\labell{sanders},\\
t&=&\lambda_0^2+\frac{(1-\lambda_0)^2}k
\;\labell{utilizzare}.
\end{eqnarray}
Equation~(\ref{utilizzare}) can be solved under the constraint
\mbox{$\lambda_1\geqslant\lambda_0$} with the result being
\begin{eqnarray}
\lambda_0=\frac{1-\sqrt{1-(k+1)(1-kt)}}{k+1}
\;\labell{ti},
\end{eqnarray}
for $t\in[1/(k+1),1/k]$. Substituting Eq.~(\ref{ti}) into the
right-hand side of Eq.~(\ref{sanders}), we obtain $S(\rho)\geqslant
{F}(t)$, with ${ F}(t)$ being the function we have plotted in
Fig.~\ref{f:plot}. Applying this result to the channel's output entropy,
we find that
\begin{eqnarray}
S(\cn_n(\rho))\geqslant { F}\left(\mbox{Tr}[(\cn_n(\rho))^2]\right)
\geqslant { F}\left(1/(2n+1)\right)
\;\labell{fine}.
\end{eqnarray}
The last inequality is derived by observing that ${ F}(t)$ is a
decreasing function of $t$ and that the minimum output R\'enyi entropy
is achieved by a vacuum-state input~\cite{renyinostro}, so that
\begin{eqnarray}
\mbox{Tr}[\cn_n(\rho)^2]\leqslant 1/(2n+1)\;.
\end{eqnarray}
Finally, (\ref{f3}) follows from~(\ref{fine}), because ${
  F}(1/(2n+1))$ coincides with the function on the right-hand side
of~(\ref{f3}).
\begin{figure}[hbt]
\begin{center}
\epsfxsize=.7
\hsize\leavevmode\epsffile{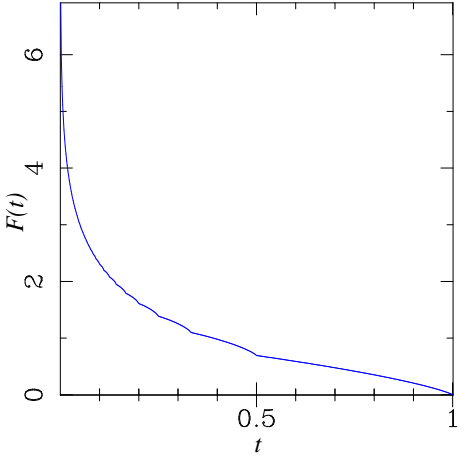}
\end{center}
\caption{Function $F(t)$ defined in Eq.~(\ref{fine}).}
\labell{f:plot}\end{figure}

\section{Derivation of lower bounds~(\ref{E}) and~(\ref{F})}\labell{s:bs}
Here we derive lower bounds~(\ref{E}) and~(\ref{F})
on the minimum output entropy $\ms(\ce_\eta^N)$.
\subsection*{Proof of lower bound~(\ref{E})}
We first prove that that (\ref{E}) applies for
$\eta= 1/k$ for all integers $k$, and then we generalize to
all $\eta$.
\begin{figure}[hbt]
\begin{center}
\epsfxsize=1.
\hsize\leavevmode\epsffile{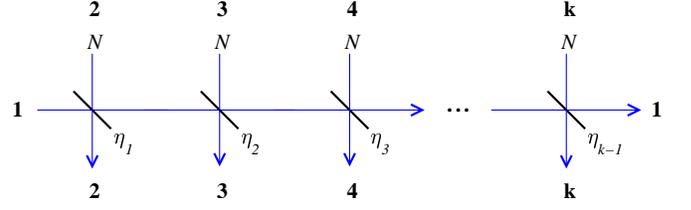}
\end{center}
\caption{Beam splitter array needed to prove (\ref{E}).
  The input and output ports are numbered so that the $j$th input is
  facing the $j$th output: the input port {\bf 1} is fed with 
  state $\rho$, the other $k-1$ ports are fed with identical thermal states
  $\tau$ of average photon number $N$. The transmissivities $\{\eta_j\}$
  are chosen in such a way that at each of the $k$ output ports 
  (indicated by arrows) we find the thermal-noise CP-map $\ce_\eta^N$
  with $\eta=1/k$. This corresponds to choosing
  \mbox{$\eta_1=(k-1)/k$}, \mbox{$\eta_2=(k-2)/(k-1)$}, $\cdots$,
  $\eta_{k-1}=1/2$.  } \labell{f:bs}\end{figure}

Consider the beam splitter array shown in Fig.~\ref{f:bs}, in which $k-1$
beam splitters of transmissivities $\eta_1,\eta_2,\ldots,\eta_{k-1}$ are
connected in series and fed with $k-1$ identical thermal states, each with
average photon number $N$. The transmissivity between the input port {\bf
1} and the $j$th output port is given by
$\epsilon_j\equiv(1-\eta_j)\eta_{j-1}\cdots\eta_1$. The beam splitters
are chosen so that $\epsilon_j = 1/k$ for all $j$, i.e., the $j$th
beam splitter has $\eta_j=(k-j)/(k-j+1)$.  For example, with $k=3$ we
have two beam splitters with transmissivities 
$\eta_1=2/3$ and $\eta_2=1/2$, respectively, so that
the transmissivity from the input
port {\bf 1} to each output port is $\epsilon_j=1/3$.  The composition
rule~(\ref{pr2}) can now be used to show that when the array is fed with a
state $\rho$, at each of the output ports (apart from an irrelevant phase
factor~\cite{nota2}) we find the state $\ce_{1/k}^N(\rho)$. The output
entropy of the joint state of all the outputs is equal to the total
entropy of the $k$ inputs, because they are connected by a unitary
transformation. For $\rho$ a pure state, this entropy is given by the
sum of the entropies of the thermal baths, i.e., $(k-1)\:g(N)$.  The
subadditivity of the von Neumann entropy~\cite{wehrl} implies that this
quantity is less than the sum of the entropies of the single outputs,
\begin{eqnarray}
k\:S(\ce_{1/k}^N(\rho))\geqslant(k-1)\:g(N)
\;\labell{qui},
\end{eqnarray}
which proves (\ref{E}) for $\eta=1/k$. The case
$\eta\leqslant 1/k$ follows immediately by using $\eta'=1/k$ in
(\ref{rel3}) and applying inequality~(\ref{qui}).  Lower
bound~(\ref{E}) for $\eta\geqslant 1/k$ is established by using
$\eta'=1/k$ in (\ref{qqq}) and again applying 
inequality~(\ref{qui}).
\subsection*{Proof of lower bound~(\ref{F})}
As in the previous case, we first prove the bound for transmissivity
$\eta=1/k$ with integer $k$ and then we extend the proof to arbitrary
$\eta$. Consider now the beam splitter array depicted in
Fig.~\ref{f:bs1}; it is the scheme considered previously with
classical noise maps added at each of the beam splitter outputs.
The composition rules Eqs.~(\ref{pr2}) and~(\ref{pr3}) show that the
output state of each port is the same~\cite{nota2}, and is given by
$\ce_{1/k}^{N+nk/(k-1)}(\rho)$. 
\begin{figure}[hbt]
\begin{center}
\epsfxsize=1.
\hsize\leavevmode\epsffile{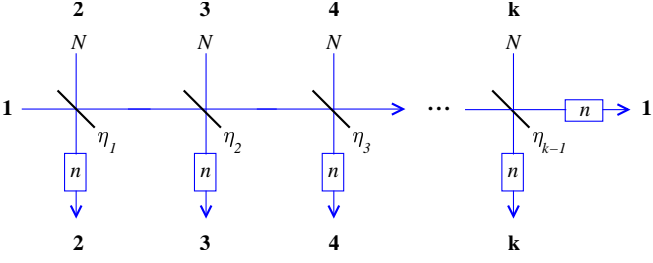}
\end{center}
\caption{Beam splitter array needed to prove (\ref{F}). Here the
  signals exiting each beam splitter encounter classical-noise channels.
  The transmissivities $\{\eta_j\}$ are chosen as in Fig.~\ref{f:bs}.}
\labell{f:bs1}\end{figure} To calculate the joint state of all the
outputs, we define $a_j$ to be the annihilation operator of the $j$th
input and
$V$ the unitary operator associated with the beam splitter array. This
operator is given by $U_1U_2\cdots U_{k-1}$, where $U_j$ is the $j$th
beam-splitter operator, defined in Eq.~(\ref{defu}), which couples the
mode $a_{j+1}$ with one of the output modes of beam
splitter $j-1$. $V$ produces the following field transformation
\begin{eqnarray}
{V}^\dag\:\vec a\:{V}=\vec a\cdot\Lambda
\;\labell{trr},
\end{eqnarray}
where $\Lambda$ is the $k\times k$ real unitary matrix with
$|\Lambda_{ij}|$ being the effective transmissivity between the $i$th
input and the $j$th output. The sign of $\Lambda_{ij}$ depends on the
reflections encountered by the field.  The sign convention that follows
from Eq.~(\ref{defu}) is that fields
propagating from left to right (see Fig.~\ref{f:bs1}) acquire a $\pi$-rad
phase shift when reflected, whereas fields propagating from
top to bottom do not suffer any phase shift when reflected. This
convention implies that
$\Lambda_{ij}$ is negative if $j>i$ and positive otherwise (e.g.,
$\Lambda_{23}=-\sqrt{(1-\eta_1)(1-\eta_2)}$, $\Lambda_{32}=0$).  Given
the input state $R=\rho\otimes\tau\otimes\cdots\otimes\tau$ ($\tau$
being a thermal state with average photon number $N$), the joint output
state of the Fig.~\ref{f:bs1} map is obtained by acting on $R$
first with $V$ and then with the classical noise maps, i.e.
\begin{eqnarray}
R'=\int {\rm d}^2\vec\mu\: 
P(\vec\mu)\:D(\vec\mu)\:{V}\:R\:{V}^\dag\:
D^\dag(\vec\mu)
\;\labell{outstate},
\end{eqnarray}
where $D(\vec\mu)\equiv\exp(\vec\mu\cdot\vec a\,^\dag-\vec
a\cdot\vec\mu\,^\dag)$ with $\vec\mu\equiv(\mu_1,\cdots,\mu_k)$, 
$\vec a\equiv(a_1,\cdots,a_k)$, and 
\begin{eqnarray}
P(\vec\mu)=\exp(-|\vec\mu|^2/n)/(\pi n)^k
\;\labell{pigrande}.
\end{eqnarray}
Using Eq.~(\ref{trr}) and performing a change of integration variables
$\vec\nu\equiv\vec\mu\cdot\Lambda^\dag$, the output state can be
written as
\begin{eqnarray}
R'&=&{V}\left[\int {\rm d}^2\vec\nu\: P(\vec\nu)\:D(\vec\nu)\:R\:
D^\dag(\vec\nu)\right]{V}^\dag\nonumber\\
&=&{V}\left[
\cn_n(\rho)\otimes\cn_n(\tau)\otimes\cdots\otimes\cn_n(\tau)\right]
{V}^\dag
\;\labell{cc}.
\end{eqnarray}
The entropy of this state is simply given by 
\begin{eqnarray}
S(R')=S(\cn_n(\rho))+(k-1)\:g(n+N)
\;\labell{cc1},
\end{eqnarray}
where we have used the fact that $\cn_n(\tau)$ is a thermal state with
$n+N$ photons on average (see Sec.~\ref{s:composition}). The subadditivity
of the von Neumann entropy implies that $S(R')$ cannot be greater than the
sum of the entropies of the individual output states, i.e.,
\begin{eqnarray}
&&k\:S(\ce_{1/k}^{N+nk/(k-1)}(\rho))
\geqslant S(\cn_n(\rho))+(k-1)\:g(n+N)
\;,\nonumber\\&&\labell{cc2}
\end{eqnarray}
which applies for any input state $\rho$.  Note that when $n=0$ we
recover inequality~(\ref{qui}), as expected.  Lower
bound~(\ref{F}), instead, derives by choosing $N=0$, so that
\begin{eqnarray}
S(\ce_{1/k}^{nk/(k-1)}(\rho))\geqslant
\frac{S(\cn_n(\rho))}{k}+\frac{k-1}k
g(n)
\;.
\;\labell{cc3}
\end{eqnarray}
As in the previous subsection, we can apply the composition
rules~(\ref{rel3}) and~(\ref{qqq}) (using $\eta' = 1/k$) to
extend the bound~(\ref{cc3}) to any value of $\eta$, obtaining 
inequality (\ref{F}).

\section{Derivation of the output-state representation (\ref{eq:numbrep})} \labell{s:numbrep}
In this final appendix, we derive the
Fock-state representation of the classical-noise channel's output state when its input is an
arbitrary pure state.  Using the Fock-state expansion of the input state,
\begin{equation}
|\psi\rangle = \sum_{n}\psi_n |n\rangle,
\end{equation}
in Eq.~(\ref{characteristic}) we immediately obtain the symmetric characteristic function at the
output of the classical-noise channel:
\begin{eqnarray}
\chi'(\mu) &=& e^{-n|\mu|^2}\langle \psi|e^{\mu a^\dagger - \mu^* a}|\psi\rangle\\ 
&=& e^{-(n+1/2)|\mu|^2}\langle \psi|e^{\mu a^\dagger}e^{- \mu^* a}|\psi\rangle \\ 
&=& \sum_{j=0}^{\infty}L_j^0(|\mu|^2)|\psi_j|^2  
+ \sum_{j=1}^{\infty}\sum_{k=0}^{j-1}\sqrt{\frac{j!}{k!}}L^{j-k}_k(|\mu|^2) \nonumber \\
&\times&[\psi_j^*\psi_k\mu^{j-k} +\psi_j\psi_k^*(-\mu)^{j-k}],
\end{eqnarray}
where $\{L^{k}_j(z)\}$ are the Laguerre polynomials.  Recovering the  output state $\rho'$ from
this characteristic function via $\int {\rm d}^2\mu\:\chi'(\mu)\:D(-\mu)/\pi$, and performing the
integration in polar coordinates, we obtain the desired Fock-state
representation of $\rho'$:
\begin{eqnarray}
\lefteqn{\langle k+l|\rho'|k\rangle = \sqrt{\frac{k!}{(k+l)!}}\sum_{j=0}^\infty
\sqrt{\frac{j!}{(j+l)!}}\,\psi_{j+l}\psi_j^*} \nonumber \\
&\times&\frac{(j+k+l)!}{j!k!}\frac{n^{j+k}}{(1+n)^{j+k+l+1}} \nonumber \\ 
&\times& F(-j,-k;-(j+k+l);1-n^{-2}),
\labell{numberrep}
\end{eqnarray}
for $k,l\ge 0$, where $F(\alpha,\beta;\gamma;z)$ is the hypergeometric function.  Note that
(\ref{numberrep}) becomes diagonal when the input is a Fock state.  In this case,
(\ref{numberrep}) can be reduced to (\ref{jacobi}) by means of a transformation formula for the
hypergeometric function and the connection between hypergeometric functions and the Jacobi
polynomials \cite{caves1}.

{\bf {Acknowledgments}}
  The authors thank P. W. Shor, B. J. Yen, H. P. Yuen and P. Zanardi for useful
  discussions.  This work was funded by the ARDA, NRO, NSF, and by ARO
  under a MURI program.


\begin{references}
\bibitem{nota} The quantity
  $q_k(\mu)\equiv\left|\langle\gamma_k|D(\mu)|\psi\rangle\right|^2/\pi$ is
  a probability distribution with respect to $\mu$ because~\cite{vourdas}
\begin{eqnarray}
\int {\rm d}^2\mu\:q_k(\mu)&=&
\int \frac{{\rm d}^2\mu}\pi\langle\gamma_k|D(\mu)|\psi\rangle\langle
\psi|D^\dag(\mu)|\gamma_k\rangle\nonumber\\&=&
\langle\gamma_k|\openone|\gamma_k\rangle
\:\mbox{Tr}[|\psi\rangle\langle\psi|]=1
\;\nonumber,
\end{eqnarray}
which can be proven using Schur's lemma since $D(\mu)$ is a unitary
irreducible representation of the Weyl-Heisenberg group.
\bibitem{nota2} Because the reflected and the transmitted photons have
  opposite phases, there is always a phase difference between the two.
  However, in the case we are considering the photons interact with a
  thermal state so that the phase difference between the two output
  ports can be compensated by a unitary transformation that does not
  change the entropy.
\end{references}
 \end{document}